\newcommand\T{\rule{0pt}{2.6ex}}       
\newcommand\B{\rule[-1.2ex]{0pt}{0pt}} 
\documentclass{aa}  

\usepackage{graphicx}
\usepackage{natbib}
\usepackage{subcaption}
\usepackage{amssymb}

\bibpunct{(}{)}{;}{a}{}{,} 
\usepackage{txfonts}
\usepackage[normalem]{ulem}
\usepackage{color}

\begin{document}

   \title{Mid-infrared interferometry of Seyfert galaxies: Challenging the Standard Model}
   \author{N. L\'opez-Gonzaga\inst{1}\and W. Jaffe\inst{1} }
   \institute{Leiden Observatory, Leiden University, P.O. Box 9513, 2300 RA Leiden,
The Netherlands  \\ \email{nlopez@strw.leidenuniv.nl}      }   
   \date{Received ;}

  
  \abstract
   {}
   {We aim to find torus models that explain the observed high-resolution mid-infrared measurements of active galactic nuclei (AGN).
   Our goal is to determine the general properties of the circumnuclear dusty environments.}
   {We used the mid-infrared interferometric data of a sample of AGNs provided by the instrument MIDI/VLTI and followed a statistical approach to compare the observed distribution of the interferometric measurements with the distributions 
   computed from clumpy torus models. 
   We mainly tested whether the diversity of Seyfert galaxies can be described using the Standard Model idea, where differences are solely due to a line-of-sight effect.
   In addition to the-line-of sight effects, we performed different realizations of the same model to include possible variations that are caused by the stochastic nature of the dusty models.}
   {We find that our entire sample of AGNs, which contains both Seyfert types, cannot be explained merely by an inclination effect and by including random variations of the clouds. 
   Instead, we find that each subset of Seyfert type can be explained by different models, where the filling factor at the inner radius seems to be the largest difference.
   For the type I objects we find that about two thirds of our objects could also be described using a dusty torus similar to the type II objects.
   For the remaining third, it was not possible to find a good description using models with high filling factors, while we found good fits with models with low filling factors.}
   {Within our model assumptions, we did not find one single set of model parameters that could simultaneously explain the mid-infrared data of all 21 AGN with line-of-sight effects and random variations alone.
   We conclude that at least two distinct cloud configurations are required to model the differences in Seyfert galaxies, with volume-filling factors differing by a factor of about 5\,--\,10. 
   A continuous transition between the two types cannot be excluded.}
   
   
   \maketitle
%
   \graphicspath{{/home/nlopez/plots/show/}
   {/home/nlopez/Documents/Papers/Paper_2/language_editor/plots/}}

\section{Introduction.}
\label{sec:intro}

Active galactic nuclei (AGN) have been extensively studied to understand the possible link between the growth of supermassive black holes (SMBHs) and the evolution of galaxies.
The main characteristic of AGNs is their extremely high luminosity.
In particular, AGNs are known to emit a large part of their energy in the form of  infrared radiation \citep[][and references therein]{1979ApJ...230...79N,1987ApJ...320..537B,1989ApJ...347...29S,1994ApJS...95....1E}. 
This infrared excess can be explained by a conversion process where a fraction of the nuclear UV and optical radiation is absorbed by circumnuclear dust at a few parsecs from the central black hole and re-emitted in the infrared regime. 
This circumnuclear dust, commonly referred to as the dusty torus, not only redistributes the emitted energy of AGNs, but sometimes also blocks our view of the nuclear engine.

According to the Standard Model for AGNs, all Seyfert galaxies are assumed to have a similar nuclear environment, consisting of an accreting supermassive black hole surrounded by ionized clouds moving at high velocities (the broad emission line region: BLR).
This nuclear engine is then surrounded by circumnuclear dust.
In its most simple form, the Standard Model predicts a bimodal distribution of the Seyfert types \citep{1993ARA&A..31..473A, 1995PASP..107..803U}: type I, for which  the nuclear engine can be directly viewed, and type II, for which the view to the central engine is blocked by dust. 
This idea is supported by the broad emission lines in the spectra of many type II sources observed in polarized light \citep[see, e.g.,][]{1985ApJ...297..621A}.

Studying the properties and morphology of the circumnuclear dust is crucial to improve our understanding of the accretion process of AGNs. 
It is unclear as yet how the gas flows into the accretion disk, but tracing the coexisting dust can help to reveal the morphology of the gas stream.
This process of transport to the inner regions is poorly understood, but is relevant for understanding the triggering and evolution of AGNs as well as the energy feedback to the host galaxy. 

High-resolution infrared images of the circumnuclear dust are expected to allow tracing the structure of these objects and determining their general properties. 
But infrared observations of the AGN environment that isolate and resolve the torus emission have been difficult to obtain. 
Early observations with the \textit{Spitzer} telescope provided studies of AGN samples \citep[see, e.g.,][]{2006AJ....132..401B}. 
Their sensitivity allowed statistical studies on a large number of detected objects, but their limited spatial resolution did not accurately isolate the AGN emission from sources of contamination, such as star-heated dust and dust in the ionization cones \citep{2000AJ....120.2904B,2001ApJ...557..637T,2005ApJ...618L..17P}.
In contrast, large ground-based MIR instruments, with their higher resolution power, can distinguish between AGN emission and star formation regions
\citep[e.g.,][]{2005A&A...438..803G, 2006ApJ...652L..83A, 2011ApJ...736...82A, 
2006A&A...457L..17H,2008A&A...479..389H,2009A&A...495..137H, 2007A&A...473..369H,
2008A&A...488...83S,2009ApJ...703..390L, 2009ApJ...702.1127R, 2011ApJ...731...92R,
2010A&A...515A..23H, 2010MNRAS.402..879R,2010A&A...511A..64V, 2012AJ....144...11M,2011A&A...536A..36A,2014MNRAS.439.1648A}.
However, in the majority of the cases the AGN emission remained unresolved, suggesting either a small size for the nuclear dusty environment or potential nonthermal contributions such as the synchrotron emission observed in the radio galaxy Centaurus A.

Mid-infrared interferometers have enabled a breakthrough by spatially resolving the compact emission of AGNs.
Several studies of individual galaxies have revealed the complexity of the nuclear dusty environment.
A few examples of these findings are that the nuclear dust environment of the Circinus galaxy shows a two-component structure consisting of a disk-like emission surrounded by an extended emission with its major axis close to the polar axis \citep{2014A&A...563A..82T};
the hot and compact dusty disk in the nucleus of NGC~1068  shows extended and diffuse emission along one side of its ionization cone \citep{2009MNRAS.394.1325R,2014A&A...565A..71L};
NGC~424 and NGC~3783 show extended thermal infrared emission with major axes close to the polar axes \citep{2012ApJ...755..149H, 2013ApJ...771...87H}.
In addition,  \citet{2011A&A...531A..99T} analyzed a sample of sources observed with interferometers and suggested a luminosity-size relation for the warm dust.
This relation was later challenged by \citet{2011A&A...536A..78K} using a sample of type I sources. It seemed similarly unlikely
in the light of results obtained using a larger sample of AGNs, the MIDI AGN Large Programme \citep{2013A&A...558A.149B}, which revealed a  diversity of complex dust morphologies on subparsec scales.
The diversity of sizes suggests that a  luminosity-size relation might not be unique for the warm dust as it is in the case of the hot dust observed in the near-infrared \citep{1987ApJ...320..537B, 2006ApJ...639...46S,2009A&A...493L..57K, 2011A&A...527A.121K, 2012A&A...541L...9W}, where the inner radius of the torus scales with the square root of the AGN luminosity.

From the theoretical point of view, much progress has been made in recent years in reproducing the infrared emission of the dusty torus with radiative transfer models. 
Early radiative transfer models of AGN dust tori were { carried out by \citet{1992ApJ...401...99P, 1993ApJ...418..673P, 1994MNRAS.268..235G}
using smooth dust distributions.}
However, such a smooth dust distribution probably does not survive in the nuclear environment of an AGN \citep{1988ApJ...329..702K}, but might instead be present in the form of clouds. 
Pioneer work from \citet{2002ApJ...570L...9N} presented a stochastic torus model with dust distributed in clumps that is capable of attenuate the strength of the silicate feature. 
More recently, many torus models, using different radiative transfer codes, techniques, and dust compositions, have been developed to obtain more efficient and accurate solutions of the radiative transfer equations and to improve the assumptions \citep{2002ApJ...570L...9N, 2005A&A...436...47D, 2006A&A...452..459H, 2008ApJ...685..147N, 2008ApJ...685..160N, 2008A&A...482...67S, 2010A&A...523A..27H, 2012MNRAS.420.2756S,2012ApJ...751...27H, 2015arXiv150804343S}.

However, all the models face one common problem: the dynamical stability of the structure and the process to maintain the required scale height are still debated.
Self-consistent models describing both the physical processes that distribute the toroidal gas and dust and the redistribution of the nuclear emission are still under development, but with promising results \citep[see, e.g.,][and references therein]{2011ApJ...741...29D,2012ApJ...758...66W,2014MNRAS.445.3878S}.

Many authors have fit the spectral energy distributions (SEDs) of Seyfert galaxies with clumpy torus models \citep[see, e.g.,][]{2009ApJ...707.1550N, 2009ApJ...705..298M, 2009ApJ...702.1127R, 2011ApJ...736...82A}, but the conclusions from these works must be examined critically.
Since the SEDs contain no direct spatial information on the torus, the results are highly degenerate; results from a comparison between clumpy and continuous models indicate that models using different assumptions and parameters can produce similar SEDs \citep{2012MNRAS.426..120F}.
We may expect the degeneracies to be partially broken if we include high-resolution interferometric observations that resolve the structures and provide direct measures of the sizes and shapes of the emission regions. 

The aim of this work is to find a family of torus models that fits the interferometric data on a set of AGNs obtained over the past decade.
We focus more on the general properties of the acceptable models than on particular characteristics provided by individual fits.
The paper is organized as follows: The main goals and motivation are explained in Sect.~\ref{sec:approach}. 
We provide information about the Seyfert sample and describe the data treatment in Sect.~\ref{sec:data}.
A brief explanation about the torus models used for this work and the method followed for our comparison is given in Sects.~\ref{sec:clumpy-torus} and~\ref{sec:method}, respectively.
The results are presented in Sect.~\ref{sec:results}. 
We discuss the general properties in Sect.~\ref{sec:discussion}.
A summary of the results is given in Sect.~\ref{sec:conclusions}.

\section{Probabilistic approach}
\label{sec:approach}

Ideally, multiwavelength high-quality infrared images of several AGNs would determine the most important dust model parameters, such as cloud sizes, disk inner radii, wavelength-dependent extensions, opening angles, and dust chemistry.
However, high-fidelity infrared imaging is not yet possible since interferometric techniques are time consuming and lack detailed phase data, and their resolution is not high enough to resolve individual clouds. 
This situation is expected to improve in a few years when the next generation of interferometers come online, for instance, GRAVITY \citep{2011Msngr.143...16E}, which will observe in K band, and MATISSE \citep{2008SPIE.7013E..70L}, which will observe in L, M, and N band.

Our ability to determine the underlying parameters of clumpy models is also limited by their stochastic nature; even when all parameters are specified, random variations in the cloud distributions may present markedly different images to the observer.

These limitations necessitate a probabilistic approach to modeling.
Our main goal is to investigate whether we can statistically reproduce the data of our whole interferometric sample by using models that have specified global properties, but where the appearance of each source is affected by unknown factors ($V_i$ with $i=1,2,3$):
1) the randomness in the positions of the clouds, 2) the inclination, and 3) the position angle of the source axis on the sky.
The stochastic arrangement of the clouds can produce different families of spectra or images of the model even when they are built with the same global parameters \citep{2006A&A...452..459H,2008A&A...482...67S}. 
The torus inclination angle is of primary importance  because it determines the chance of viewing directly (low inclinations) or indirectly (high inclinations) heated clouds.
The position angle is an important unknown when only limited interferometric baselines are available.

We aim to find global properties that the AGNs might have in common and to test with these the existence of any overall unifying model of AGNs.
Our procedure is to search for a model that explains all the observations on a statistical basis.
If this fails (as it does), we consider the possibility that the model parameters may vary from object to object, or for certain classes of objects.
This is the case if our models can fit each galaxy or class individually, but not all of them for one parameter set.

\section{Observational data}
\label{sec:data}

\small
\begin{table}
   \centering
   \small
   \begin{tabular}{c c c c c c c c c}
      \hline\hline
      Source &  D     & $z$ & Type & L$_{IR}$ & L$_{xray}$  \T \\
        &  [Mpc] &  &   &  &  &  \\
      \hline \hline
      I Zw 1            & 222  & 0.0589 & Sy 1 & 44.9 & 43.7\T \\
      NGC 424           & 44.7 & 0.0110 & Sy 2 & 43.6 & 43.8\\
      NGC 1068          & 14.4 & 0.0038 & Sy 2 & 44.0 & 43.6\\
      NGC 1365          & 18.1 & 0.0055 & Sy 1 & 42.5 & 42.1\\
      IRAS 05189-2524   & 167  & 0.0426 & Sy 2 & 44.6 & 43.7\\
      H 0557-385        & 135  & 0.0339 & Sy 1 & 44.4 & 43.8\\
      IRAS 09149-6206   & 222  & 0.0579 & Sy 1 & 44.9 & 44.0\\
      MCG-05-23-16      & 38.8 & 0.0085 & Sy 2 & 43.5 & 43.3\\
      Mrk 1239          & 84.5 & 0.0200 & Sy 1 & 44.0 & 43.3 \\
      NGC 3281          & 47.6 & 0.0107 & Sy 2 & 43.4 & 43.2\\
      NGC 3783          & 43.8 & 0.0097 & Sy 1 & 43.7 & 43.2\\
      NGC 4151          & 16.9 & 0.0033 & Sy 1 & 43.0 & 42.5\\
      NGC 4507          & 51.7 & 0.0118 & Sy 2 & 43.7 & 43.2\\
      NGC 4593          & 41.2 & 0.0090 & Sy 1 & 43.1 & 42.9\\
      ESO 323-77        & 64.2 & 0.0150 & Sy 1 & 43.7 & 42.8\\
      IRAS 13349+2438   & 393  & 0.1076 & Sy 1 & 45.5 & 43.9 \\
      IC 4329 A         & 68.3 & 0.0161 & Sy 1 & 44.2 & 43.9\\
      Circinus          & 4.2  & 0.0014 & Sy 2 & 42.7 & 42.3\\
      NGC 5506          & 28.7 & 0.0062 & Sy 2 & 43.4 & 43.1\\
      NGC 5995          & 102  & 0.0252 & Sy 2 & 44.1 & 43.5\\
      NGC 7469          & 60.9 & 0.0163 & Sy 1 & 43.9 & 43.2\B \\
      \hline \hline
   \end{tabular}
   \normalsize
   \caption{Source properties.  {\bf Notes.}
   {\it D}: angular-size distance derived from redshift, except for Circinus and NGC~1068; 
   {\it z}: Redshift (from NED);
   {\it Type:} AGN classification from SIMBAD;
   {\it L$_{IR}$}: The 12\,$\mu$m infrared luminosity is given as $\log(L_{MIR}/erg\cdot s)$ and 
   the values were obtained from \citet{2013A&A...558A.149B}, uncertainties are typically lower than 5\%;
   {\it L$_{xray}$}: The absorption corrected 2\,--\,10\,keV X-ray AGN luminosity given as  $\log(L_{x}/erg\cdot s)$. 
   The values were collected from \citet{2015MNRAS.454..766A}}
   \label{tab:data}
\end{table}
\normalsize

\subsection{Infrared data} 
\label{subsec:ir-data}

Our sample consists of 21 Seyfert galaxies observed with the MID-infrared Interferometric instrument \citep[MIDI\footnote{The instrument MIDI is a two-telescope Michelson-type beam combiner with an operational spectral range in the atmospheric $N$ band ($\lambda \sim 8\,--\,13\mu$m)}][]{2003Ap&SS.286...73L} at the European Southern Observatory's (ESO's) Very Large Telescope. 
This flux-limited sample was published by \citep{2013A&A...558A.149B}, who required sources with a flux higher than 300\,mJy at $\lambda\sim$12\,$\mu$m in high-resolution single-aperture observations.
For more specific information about the reduction process and observation strategy we refer to \citet{2013A&A...558A.149B}.
The original set also includes data for the quasar 3C273 and the radio source Cen~A, but we omit these two sources as their nuclear mid-infrared emission { may not originate in the same way as in Seyfert galaxies.}
For example, the nuclear mid-infrared flux of Cen~A is dominated by unresolved emission from a synchrotron source, while the contribution of the thermal emission of the dust only represents about 40\,\% of the total emission at $12\,\mu$m \citep{2007A&A...471..453M}.
We exclude the quasar 3C273 because the mid-infrared environment of high-luminosity objects might differ from that of low-luminosity objects (Seyfert galaxies) sources \citep[see, e.g.,][]{2015ApJ...807..129S}.  

Each source was observed with pairs of 8\,m unit telescopes (UTs), and Circinus and NGC~1068 were additionally observed with pairs of 1.8\,m auxiliary telescopes (ATs), in at least three different baseline configurations. 
The main observable of the instrument MIDI is the {\it \textup{correlated flux}}, which can be seen as the measured fraction of the total flux that is coherent for a particular ($u,v$) point\footnote{A ($u,v$) point can be defined as the coordinates, for a given projected baseline and a position angle, in the Fourier-transform space of the angular distribution of the source on the sky.}.

To capture the shape of each interferometric spectrum and to reduce the computational time, we used three different wavelengths in our fits.
We took the average values at (8.5 $\pm$ 0.2)\,$\mu$m, (10 $\pm$ 0.2)\,$\mu$m and (12 $\pm$ 0.2)\,$\mu$m rest frame, whereby we include information about the slope and the amplitude of the silicate feature.

\begin{figure}
   \centering
   \includegraphics[width=\hsize]{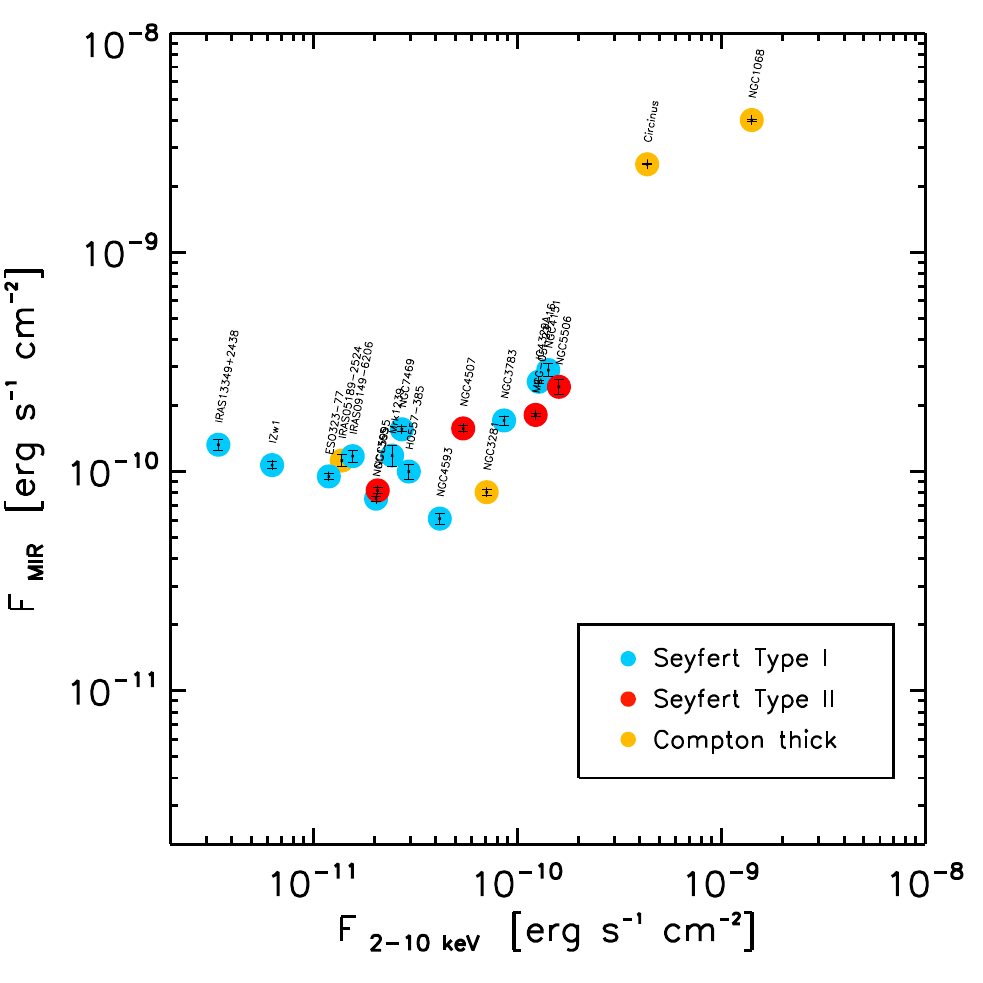}
   \caption{ Absorption corrected 2\,--\,10\,keV X-ray  fluxes versus nuclear 12\,$\mu$m fluxes. }
   \label{fig:xray}
\end{figure}
   
\section{Clumpy torus models}
\label{sec:clumpy-torus}

Since we cannot create images from our mid-infrared interferometric data, we need to make use of models to interpret our observations.
The dusty cloud models used for this work are based on the approach followed before by \citet{2008A&A...482...67S}. 
In this section we briefly explain some of the general aspects of the models, but for more details we refer to Appendix A.
These models are built with dense spherical dusty clouds distributed randomly throughout a defined volume. 
The temperature distributions within these cloud arrangements are quite complex and need to be solved numerically by using radiative transfer codes.
The overall dust temperature distribution of the dust and the scaling properties of the torus are essentially determined by the strength of the heating source and the fraction of the UV emission that the dust clouds intercept. 

The models used for this work are characterized by the following parameters ($P_i$):
1) The total average {\it \textup{optical depth}} at 9.7\,$\mu$m along the equator,
2) the {\it \textup{opening angle}} defining the dust-free cone,
3) the local fractional volume occupied by the dusty clouds at 1\,pc given by the \textup{{\it \textup{filling factor}}},
4) the {\it \textup{radial extension}} of the dusty torus defining the outer radius,
and 5) the \textup{\textup{{\it \textup{density profile index}}}} $\alpha$  that defines the radial density profile of the clouds.
The modifications to the original model presented by \citet{2008A&A...482...67S} are as follows.
\begin{itemize}
   \item Isotropic emission of the central source.
   We omit the $|\cos(\theta)|$ law profile emission of the original model as there is no evidence for a strong anisotropy in the mid-infrared x-ray relation.
   \item We define our filling factor at the inner 1\,pc region instead of taking the whole volume space. 
   N-band fluxes are sensitive to dust with a temperature near 300\,K, and for the nuclear luminosities $L_{UV}$ used in our modeling, most of the dust at this temperature is found at a radius of $\sim 1$\,pc.
\end{itemize}
 
We used the radiative transfer code RADMC-3D\footnote{http://www.ita.uni-heidelberg.de/~dullemond/software/radmc-3d/} to compute the temperature and the surface  brightness distributions of the dusty torus.
First the temperature of the system was computed by sending out photon packages using a Monte Carlo approach. 
Anisotropic scattering was treated using the Henyey-Greenstein approximate formula (\cite{1941ApJ....93...70H}.
After computing the temperature of the dust grains, we used the included ray tracer to obtain the surface brightness maps at the required wavelengths.
We computed high-resolution model images for different lines of sight (a given $\phi$ and $\theta$ angle in the coordinate system of the model) at three different 
wavelengths, 8.5\,$\mu$m, 10.0\,$\mu$m, and 12.0\,$\mu$m.
To determine the corresponding Seyfert type (I or II) of the images along a line of sight, we took the respective value of the optical depth in the visual $\tau_V$ and classified them as type I if $\tau_V < 1$ and type II if $\tau_V > 1$, that is, type I if there is a direct view of the nucleus and type II if the nucleus is obscured.
Finally, to obtain the  correlated fluxes, we applied a discrete fast Fourier transform to each image.

For every parameter set, we computed at least ten different realizations of the model to estimate the variations that are due to the position of the clouds.
For every realization we extracted the images along ten different lines of sight corresponding to type I objects and also ten lines of sight where the nucleus is obscured (type II objects).

\small
\begin{table}
\centering

\small
\begin{tabular}{l c }
\hline\hline
\multicolumn{2}{c}{INPUT PARAMETERS} \T\\
Parameter & Values   \T \\
\hline
\hline
Bolometric luminosity accretion disk $(L_{disk})$       &  $1.2 \times 10^{11}\,L_\sun$ \T \\
Inner radius of the torus $(R_{in})$                    &  0.4\,pc\B \\
Constant of clump size distribution $(a_0)$             &  0.2\,pc\B \\
Radial profile density exponent $(\alpha)$              &  $-2, -1.5, -1, -0.5, 0$ \B \\
Radial extension                                        &  25, 50, 75, 100 $R_{in}$ \B \\
Half opening angle $(\theta_{open})$                    &   30$\,^\circ$, 45$\,^\circ$, 60$\,^\circ$ \B \\
Total average $\tau_{9.7}$ along the equator $(\langle\tau_{9.7}\rangle_\phi)$ &   1, 2, 4, 8, 16 \B \\
Filling factor at inner rim                             & 0.4, 1.4, 5.3, 20, 40\,\% \B \\
Number of realizations                                  & 10 \B \\
Lines of sight per realization                          & \\
Type I:                                                 & 10 \B \\
Type II:                                                & 10 \B \\
Distribution of inclination angles                      & Uniform in a sphere\B \\
\hline
\hline
 \end{tabular}
 \normalsize
 \caption{Input parameters. Values of the parameters used as input to build 
 the clumpy torus models. For a full description of how the torus models are 
constructed see Appendix~A.}
\label{tab:parameters}
\end{table}
\normalsize

\subsection{Luminosity rescaling}
\label{subsec:rescaling}

The luminosity of the central engine obviously is a key parameter in determining the appearance of the source. 
To match our model images with the observational data for any particular source, we required an accurate estimate of the nuclear UV luminosity to scale the size of the observed objects with the size of the model images.
Since it is usually not possible to directly measure the UV emission of the accretion disk, we examine here one of the commonly used tracers for the UV luminosity: the absorption-corrected 2--10\,keV X-ray luminosity.

Several studies \citep[see, e.g.,][]{2004A&A...418..465L, 2008A&A...479..389H, 2009A&A...502..457G, 2009ApJ...703..390L} have reported a tight correlation between absorption-corrected 2\,--\,10\,keV X-ray and mid-infrared luminosities for Seyfert galaxies, which  has been interpreted as a direct connection between the luminosity of the accretion disk and the luminosity of the torus.
In Fig.~\ref{fig:xray} we show the mid-infrared fluxes from \citet{2013A&A...558A.149B} and the absorption-corrected 2--10\,keV X-ray fluxes from \citet{2015MNRAS.454..766A}.
We used fluxes instead of  luminosities to avoid false correlations induced by the spread in redshifts.
The correlation is unclear and the X-ray flux is spread over about one decade for sources with essentially the same MIR flux.

Because the relation of the X-ray and UV as well as the significant scatter in the MIR-Xray is unclear, we decided to avoid using $L_X$ as a proxy for $L_{UV}$. 
Instead we assumed a nominal nuclear isotropic heating luminosity $L_m$ as part of the modeling process and adjusted its value for each model so that the single-aperture 12\,$\mu$m predicted by the model matches the observed value.  
The observed single-aperture fluxes are in general accurately measured.  
This $L_m$ was used to rescale the model sizes and fluxes, as described below.

The nominal luminosity $L_m$ is the energy emitted from the nucleus that then iluminates the clouds and generates the 12\,$\mu$m emission.
This luminosity is effectively the same as $L_{UV}$, although it might differ slightly if the dust at the inner radius is not modeled accurately. 
Although we cannot strictly check the accuracy of this nominal luminosity because we lack NIR measurements, we expect the deviations to be only mild as the hot emission is treated consistently in our models. 
Thus any possible deviation from the true $L_{UV}$ might occur if a completely different prescription for the ensemble of clouds were used in the enviremissiononment close to the sublimation radius.

The images described in the previous section were computed for a nominal model nuclear UV luminosity $L_m$ (1.2$\times 10^{11}\,L_\sun$) at a nominal model distance $D_m$. 
These must be compared to the mid-infrared observations of sources at an actual distance $D_s$ computed from the redshift and actual nuclear luminosity $L_s$, which is assumed to be unknown

For this comparison, we mathematically  moved the model to $D_s$ and then adjusted $L_m$ until the total infrared continuum emission toward the observer at $\lambda=12\,\mu$m equaled the observed 12\,$\mu$m single-aperture flux.
This adjustment was calculated for each model realization, including the cloud distribution and the inclination angle $\theta$, because these factors affect the fraction of the nuclear luminosity converted from UV into infrared and then projected toward the observer, that is, the observed UV-IR efficiency, $\eta_{UV-IR}$.

Adjusting the nuclear luminosity would a priori involve recalculating the radiation transferred through the cloud distribution for each realization.
Fortunately, scaling relations in the radiation transfer obviate this computationally expensive step. 
Assuming that the grain size distribution remains constant, we expect the inner dust sublimation radius $r_{in}$ to scale as $L_s^{1/2}$ because the temperature of dust grains exposed directly to nuclear UV should only depend on the flux, $L_s/r_{in}^2$. 
So we may intuitively expect a source with a given $L_s$ to resemble one with $L_m$ , but all emitted luminosities are scaled by $L_s/L_m$ and all dimensions scaled
by $(L_s/L_m)^{1/2}$.
We directly tested this scaling relation with the RADMC-3D models over variations of a factor of 10 in $L_m$ and found it to apply with high accuracy, even for emission from regions not directly heated by the nucleus.
In other words, the spatial distribution of the infrared radiation at all wavelengths considered here scales directly with $r_{in}$. 

For a full description of our procedure we refer to Appendix~\ref{sec:scaling}.

\section{Description of the method}
\label{sec:method}

\subsection{Stochastic modeling}

In Section \ref{sec:approach} we explained that with our data, studies of individual sources may  not determine uniquely the parameters $P_i$ underlying the stochastic models.
A statistical method dealing with the entire dataset may give better insight into these parameters.

We sought a statistical method that is robust and relatively familiar, so that bad fits can be easily diagnosed.
The second criterion suggests a variant of the $\chi^2$ method. 
Several difficulties immediately arose.
First, the interferometric dataset is very inhomogeneous; measurements were made of galaxies of different luminosities, at different distances, position angles, and baselines.
Second, some of the measures are highly correlated with respect to the stochastic variables $V_i$.
Finally, the actual selection criteria of the sample are also quite inhomogeneous.

We circumvented the first two problems by using the information provided by the models to find transformations that convert the measurements $CF_{uv,n}$ to new uncorrelated, zero-mean, unit-variance variables $cf_{uv,n}$.
For each model we produced a large number of realizations of the stochastic variables $V_i$: source orientation, inclination, and cloud positions.
For each galaxy the individual measurements $CF_{uv,n}$, that is, the correlated fluxes at each $(u,v)$ position, were simulated for each of the model realizations after adjusting for the source luminosity described in Sect. \ref{subsec:rescaling}.
These simulations produced a probability distribution of simulated measurements $CF^{model}_{uv,n}$ for the galaxy that were then convolved with the distribution of noise estimates from the actual measurements $CF_{uv,n}$.
If the model is correct, the true data values should then lie within the most likely parts of the distributions (65\,\% of the distribution for a Gaussian-like distribution).
A very poor model can be rejected at this phase if the individual measurements $CF_{uv,n}$ lie outside the predicted ranges.
But models can also be rejected if the total set of data, per galaxy or group of galaxies, is unlikely, and for this we have to consider the expected correlations between the measurements.

The distributions are characterized by their means and (co)-variances.  
For a given model we now constructed for each galaxy a new set of variables $cf_{uv,n}$ from the original measurements $CF_{uv,n}$ by first subtracting the mean expectation values predicted from the models and then computing linear combinations of the measurements that diagonalize the cross-correlation matrix to unit values. 
These new variables therefore have zero mean, unit variance, and zero cross-correlation if the model is correct.
In this way we can test for any single galaxy, or any set of galaxies, the acceptability of the model by summing the squares of the transformed variables $cf_{uv,n}$  and comparing the total with that expected from the sum of the same number of normal Gaussian variables, that is, a $\chi$-squared distribution with the given number of degrees of freedom.
We note that models can be rejected if the squared sum is either too large (measurements do not look like the model) or too small (model predicts variances that are larger than the measured values).  

\subsection{Selection effects}

We now considered the inhomogeneous selection criteria.
The large program sample was chosen from well-known relatively nearby southern Seyfert galaxies, whose nuclear single-aperture N-band fluxes were above 300\,mJy.  
When we test whether a specific model could account for a single $(u,v)$ measurement or for all the measurements on one galaxy, this selection process is not critical to the interpretation. 
In this case, we only gauge whether there is some mildly probable cloud configuration that matches that data.
When instead we test whether the data from all sample galaxies, or a subgroup of these galaxies, can be explained by a single model, we have to consider the selection effects.
The data distributions calculated above were found by assuming that all the stochastic variables are uniformly distributed.  
The selection process may skew these distributions.  
For example, if a cloud distribution tends to extinguish N-band emission in the equatorial plane, galaxies with dust structures viewed edge-on will be less likely to meet the 300\,mJy limit.
This would contradict the assumption of a random distribution of inclination angles.

To account for this, we considered the efficiency $\eta_{UV-IR}$ of converting nuclear UV emission into mid-infrared emission directed toward the observer.  
Each model cloud realization yields a different calculable value for this efficiency.  
Low-efficiency realizations require a higher and therefore less probable nuclear luminosity for the mid-infrared flux to exceed the survey limit $S_{IR}$.
Therefore we modeled the effect of the mid-infrared flux selection on the model distributions by reweighting each stochastic realization proportional to the probability that the
nuclear luminosity $L_{nuc}$ exceeds $4\pi D_s^2 S_{IR}/\eta_{UV-IR}$.  
Hard X-ray surveys of Seyfert nuclei \citep{2010A&A...509A..38G} indicate that the  integral luminosity function, that is, the probability that the luminosity exceeds a specified value $L_{nuc}$, scales approximately as $L_{nuc}^{-\gamma}$ with $\gamma\simeq 1$.
Thus we can model the effect of the flux selection on the observed distributions by reweighting each realization in proportion to $\eta_{UV-IR}^{+\gamma}\simeq \eta_{UV-IR}^{+1}$.

Similarly, we introduced a reweighting to model the selection of the galaxies as Seyfert AGNs in the first place.
The principle Seyfert classifications depend on the escape of hard UV photons from the hot accretion disk to the narrow line region (NLR), which is well outside our modeling
region, where they induce high-excitation ionization.
We modeled this effect by calculating for each realization the UV-escape efficiency $\eta_{esc}$ for UV photons to escape the cloud regions and by reweighting the realization proportional to $\eta_{esc}^\gamma$.

The effects of these reweighting schemes on the best-fitting model parameters are described in more detail below.

\small
\begin{table}
   \centering
   \small
   \begin{tabular}{l c c }
      \hline
      \hline
      \multicolumn{3}{c}{Best-fit values}\T \\
      \hline
     Type I parameters                          & Acceptable area    & Best fit\T \\
\hline
Radial profile density exponent                 &  $\leq$ $-1.5$                 & $-1.5$ \T \\
Radial extension  [$r_s$]                       &   Unconstrained         & 50\\
Opening angle [Deg]                             &   Unconstrained         & 45\\
Total average $\tau_{9.7}$                      &  $\geq$ 8                     & 16\\
Filling factor at inner rim [\%]                &  [ 0.4 - 1.4 ]       & 0.4\\
\hline
     Type II parameters                                 & Acceptable area    & Best fit\T \\
\hline
Radial profile density exponent                 &   $\leq$ $-1$          &$-2$ \T \\
Radial extension  [$r_s$]                       &  Unconstrained & 100 \\
Opening angle [Deg]                             &   [45 - 60]          & 60 \\
Total average $\tau_{9.7}$                      &  $\geq$ 8              &  16\\
Filling factor at inner rim [\%]                &  $\geq$ 5           & 20\\
\hline
   \end{tabular}
   \normalsize
   \caption{Best-fit parameters.
   Range of the acceptable values and best-fit solution for each AGN subsample. 
   These acceptable solutions were obtained independently for each subsample.}
   \label{tab:parameters}
\end{table}
\normalsize

\begin{figure*}
   \centering
   \begin{subfigure}{0.48\hsize}
      \includegraphics[width=0.95\hsize]{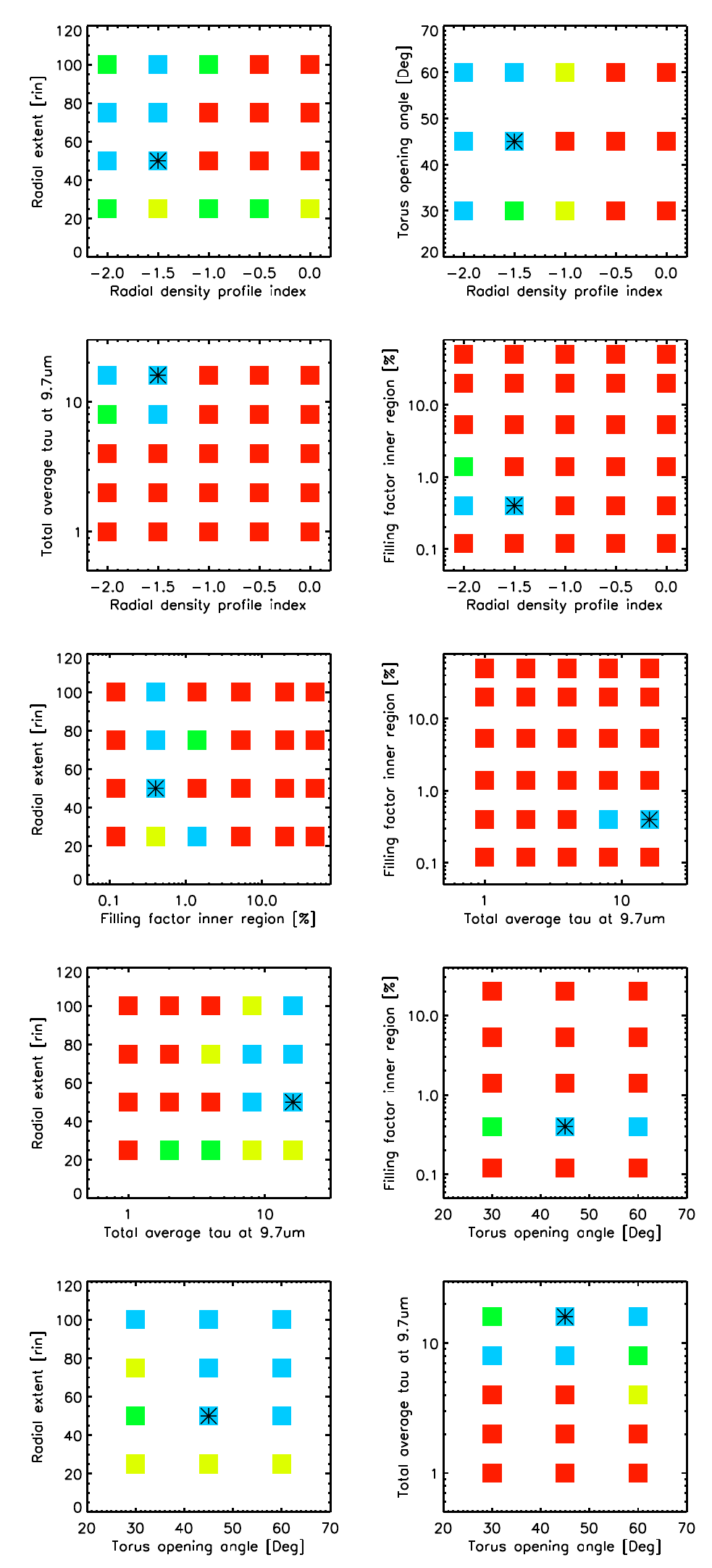}
      \caption{Type I objects}
      \label{fig:maps1}
   \end{subfigure}
   \vline
   \begin{subfigure}{0.48\hsize}
      \includegraphics[width=0.95\hsize]{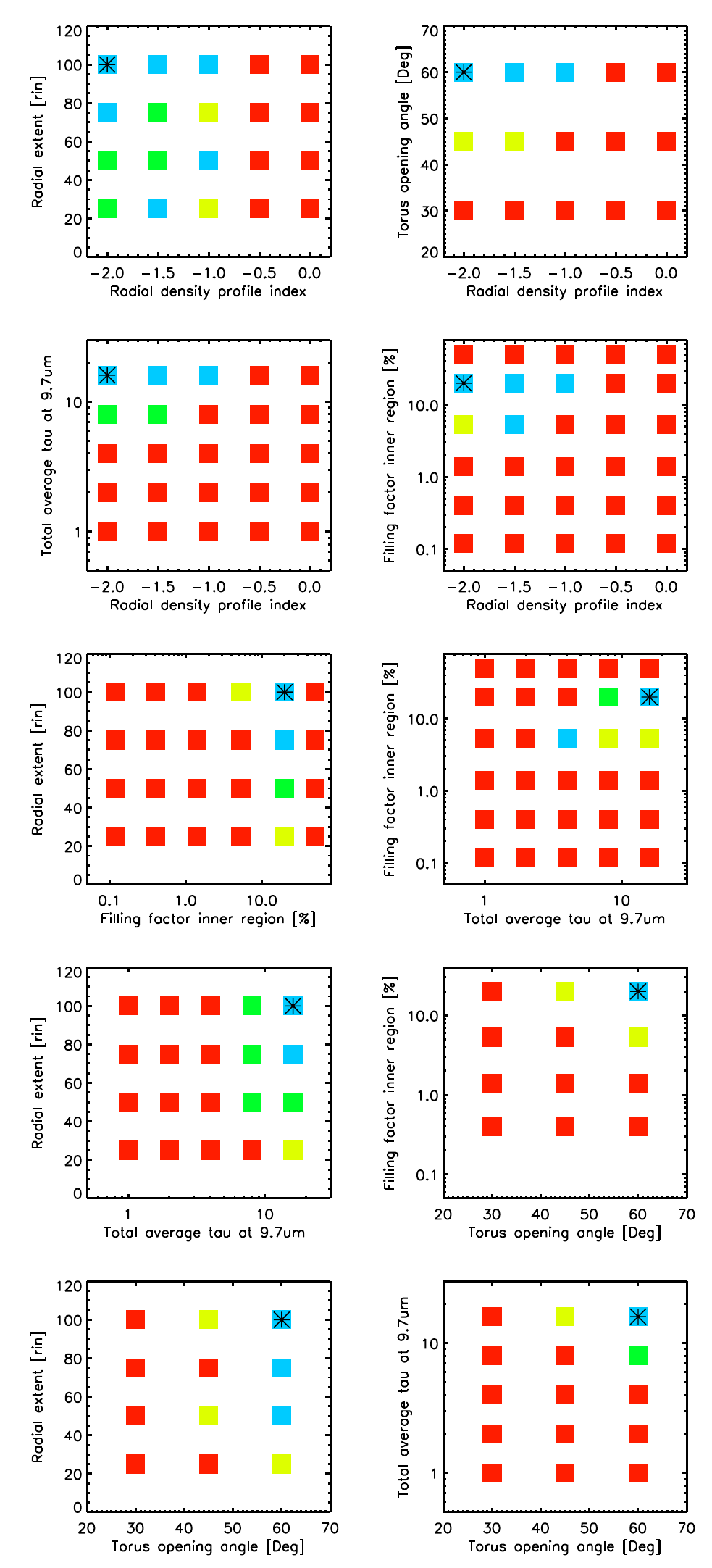}
      \caption{Type II objects}
      \label{fig:maps2}
   \end{subfigure}
   \caption{Discrete maps showing the level of acceptance around the best-fit solution for type I sources (first and second column) and type II sources (second and third column). 
   In every panel, three of the five parameters of the best solution are kept fixed, while the two parameters shown in the labels of each plot are explored. 
   The best-fit solution is shown with an asterisk.
   The color of the squares indicates the acceptance value of the parameters based on a chi-square test.
   The blue squares indicate the probability of the chi-squared value using the equivalent percentages of a Gaussian distribution at the 1-sigma level (68\,\%), green the probability at 95\,\% confidence, yellow up to 99.5\,\% and red above 99.5\,\%.}
\end{figure*}

\section{Results}
\label{sec:results}

It is very time consuming to computate the temperature profile and the respective images of every realization, therefore we explored the parameter space using a discrete set of values. 
The values taken for each parameter are shown in  the top section of Table~\ref{tab:parameters}, together with other input parameters. 
To account for a bias that is due to the detection limit of the sample, our results were obtained using a reweight, as stated in Sect. 5.2, with a value of $\gamma = 1$.

\subsection{Full sample}
\label{subsec:full-sample}

We first analyzed our entire sample containing Seyfert type I and type IIs together.
We searched for the best combination of parameters $P_i$ that statistically describe our sample.
In all our mapping space we did not find any set of parameters $P_i$ that produces models consistent with our entire sample of AGNs.
Within our range of parameters, this result suggests that our sample is not consistent with the idea that their observed differences should only be attributed to a line-of-sight effect; this is consistent with the result of \citet{2013A&A...558A.149B}. 

Our entire sample cannot be reproduced statistically by a single set of model parameters $P_i$, while each individual galaxy can be fit by its own set of parameters $P_{i,n}$. This suggests two possible cases.
First, AGNs cannot be explained with one fixed set of parameters $P_i$, but instead we need a broad range of parameters $P_i$.
Alternatively, there are major subgroups within the sample, each of which can be fit with its own set of parameters $P_i$. 
When searching for the best set of parameters, we observed that occasionally the type II objects and some of the type I objects were consistent with each other, but a significant fraction of type Is seemed to be poorly fit. 
This motivated us to investigate both types independently to search for their best-fit models.
A reasonable set of parameters that describes our subsets would allow us to explain why we failed to fit the two groups together.

\subsection{Type I}
\label{subsec:type-1s}
We continued our search in the parameter space using the type I set, that is, sources where our view to the nucleus is not blocked by the dust. 
In the models, this means taking line of sights that penetrate the dust-free volume inside of the opening angle and lines of sight at high inclinations that by chance do not encounter clouds along its path.
For this set of objects we did find combinations of model parameters that produced a distribution of correlated fluxes that is compatible in a probabilistic sense with the observational data.
The range of model parameters that shows a best fit with the data are listed in Table~\ref{tab:parameters}.
Additionally, we plot in Fig.~\ref{fig:maps1} discrete maps showing the behavior of the level of acceptance when we let two parameters change freely around the best solution.
We display these confidence levels as color-coded plots for different pairs of input parameters. 
This allows the viewer to decide quickly whether the best-fitting parameters are correlated, or in other words, whether particular combinations of parameters are better constrained than the individual parameters themselves.

We observe from Fig.~\ref{fig:maps1} that for the type I subset the best-constrained parameters are the volume-filling factor, the radial density profile index, and the optical depth. Only model spatial filling factors at 1\,pc radius between 0.4\,\% and 1.4\,\% fit the type I observations at better than the 3$\sigma$ level.
As we explain in more detail in the discussion section, the low percentages for the filling factor are necessary to produce a diverse family of spectra and sources with multiple sizes without using different parameters $P_i$ for every object.
Because the clouds are somewhat larger than in other available models \citep{2006A&A...452..459H,2008ApJ...685..147N}, realizations in our models with low filling factors and steep radial density profiles have a very limited number of total individual clouds, between 5\,--\,10 clouds on average.

We also obtain a good estimate for the total radial optical depth at 9.7\,$\mu$m. 
This parameter must be on the order of or higher than $\tau_{9.7\,\mu m}=8,$ corresponding to a value in the optical of $\tau_{0.5\,\mu m} \gtrsim 75$. 
Lower values for the optical depth all yield fits equivalent to 4$\sigma$ or worse for a normal distribution. 
A combination of high optical depths and inclination effects allows a reduction of the silicate feature. 
In this case the shadowing effect explained by \citet{2008A&A...482...67S} might not be too strong because there are only a few clouds.

With the low filling factor of the type Is, we anticipate that the index of the radial density profile for the clouds may be poorly constrained. 
The low number of clouds may not allow an accurate determination of the density profile.
Still, we observe in Fig.~\ref{fig:maps1} that only steep radial distributions, with an index lower than -1, agree with our observations, meaning that clouds are more likely to be found at close or intermediate distances (a few tenths of the sublimation radius). 
Sets of models with a low filling factor at the inner radius but flatter distributions instead produce an excess of cold emission because a many clouds could exist at large distances, while the number of clouds at the smallest distances can be kept low.

The maximum radial extension is poorly constrained in our models because we lack long-wavelength infrared data. 
In all the plots that show the radial extension the fits are good for all the possible values. 
Because the best-fit parameters have a steep radial cloud distribution, only {a very limited number of} clouds could exist at larger radius.
Clouds at large distances will be dominated by cold emission (< 100\,K) { and should be detected at (sub-)milimiter wavelengths.} 
The opening angle for a limited number of clouds does not make much sense as the distribution of clouds along the azimuthal direction produces similar results for different opening angles. 
We observe in the two lower plots of Fig.~\ref{fig:maps1} that the opening angle is essentially good in the range used for our search, 30\,--\,60 degrees. 
This behavior is expected when the number of clouds is low. 

\begin{figure}
   \centering
   \begin{minipage}{0.5\hsize}
      \centering
      \includegraphics[width=\hsize]{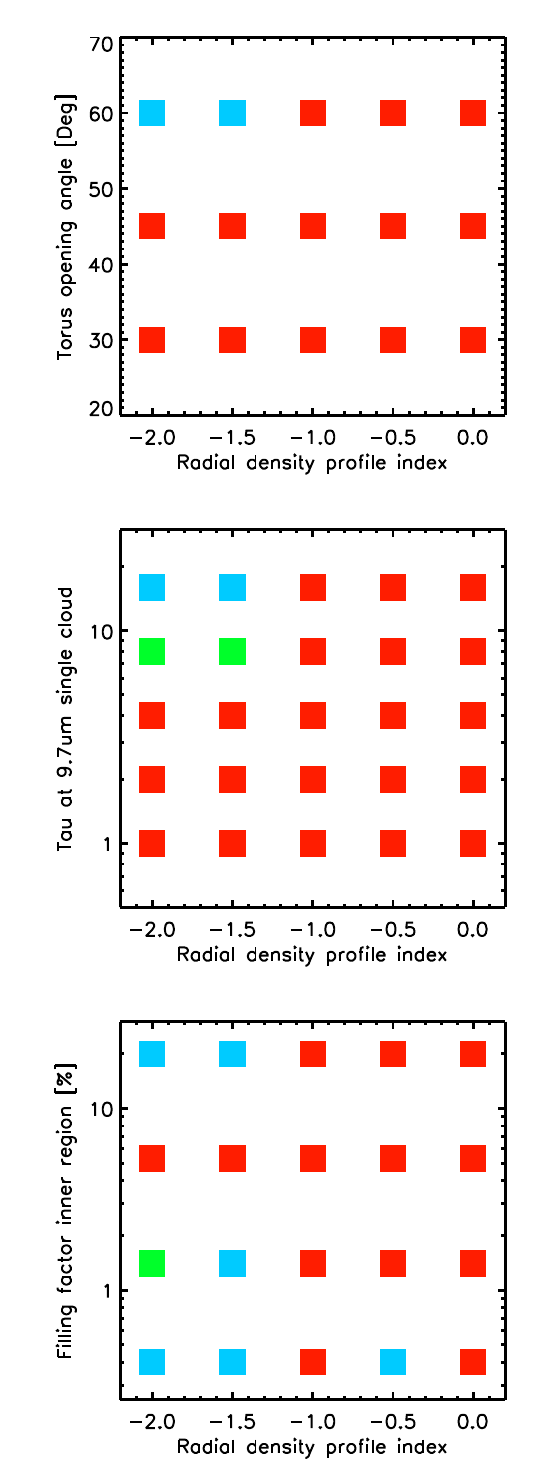}
   \end{minipage}%
   \begin{minipage}{0.5\hsize}
      \centering
      \includegraphics[width=\hsize]{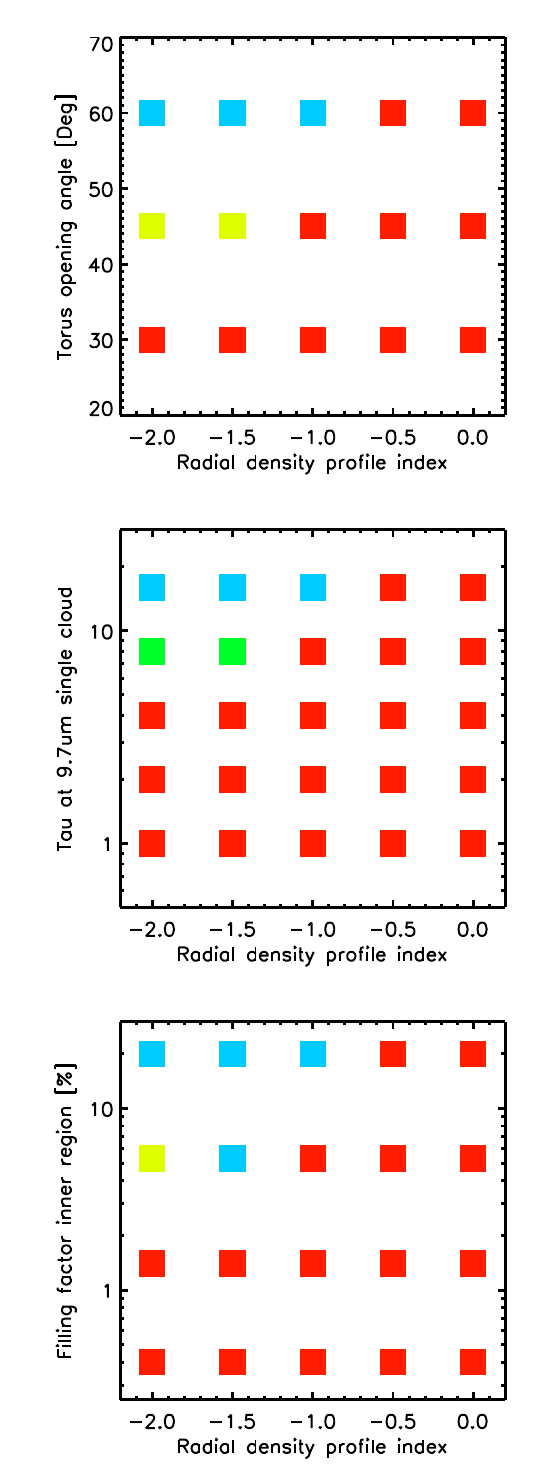}
   \end{minipage}
   \caption{Comparison of the discrete maps for models of type IIs using ({\it left}) no  luminosity reweighting and ({\it right}) a more realistic reweighting with $\gamma=1$.}
   \label{fig:weigamma}
\end{figure}

\subsection{Type II}
\label{subsec:type-2s}

After finding the best-fit parameters for the type I objects, we searched for the best-fit parameters for the type II objects to see if the deviate from the type I sample.
Since our models are wedge-like structures, the type II lines of sight are confined within a region outside the opening angle. 

We found several sets of parameters $P_i$ that reproduce the type II observations.
The range of best-fit parameters found for this subset are shown in  Table~\ref{tab:parameters}, and
variations of the level of acceptance when changing two parameters around the best-fit solution can be seen in Fig.~\ref{fig:maps2}.

For type II Seyfert galaxies, the radial slope of cloud distribution is steep, similar to the type I sources. The acceptable index for the density profile lies between $-2$ and $-1$. Again, because of this steep slope and the lack of long-wavelength infrared data, the maximum radial extension of the torus is poorly defined.

The main difference between Figs.~\ref{fig:maps1} and~\ref{fig:maps2} is the range of acceptable values for the filling factor. 
For the type II sources, the number of clouds at the inner regions is significantly larger for than the type I sources. 
The acceptable filling factors for the type II sources are larger than $\sim$5\,\%.
This means that the cloud-filled volume near 1\,pc is a factor of 5 or higher than that in the type Is. 
This suggests an important intrinsic difference between types I and II.

The average optical depth throughout the whole disk at 9.7\,$\mu$m is similar for the type II models to that of the type I models. 
Any optical depth value above 8 gives reasonable fits.
Increasing the value of $\tau$ beyond $\sim 8$ makes no difference; by this time, all the N-band photons have been absorbed and converted to even longer wavelengths.

A higher filling factor at the inner radius for type IIs and similar density profile index with respect to type Is means that type II objects have a larger total number of clouds.
With similar values of the optical depth for both types, but type IIs having a higher total number of clouds means that for type IIs it is more likely to observe dusty clouds along the line of sight. The fewer clouds in the type I model result in a longer line of sight  and ($u,v$) variations for an individual type I with respect to a typical type II. 

The best value for the opening angle lies near 60\,$^\circ$. Models with opening angles of 45\,$^\circ$ are in the 2 or 3-sigma region, those with opening angles of 30\,$^\circ$ are quite unlikely since they are in the 4-sigma area.
For small opening angles the images along obscured lines of sight of these models will look more or less round, while models with large opening angles essentially produce flat disks.
A study of the elongations in the Large Program sources suggests an intrinsic ratio of 1:2 \citep{2016arXiv160205592L}.
Therefore, a roundish model might not be a good representation of the dusty structure of sources such as NGC~1068 and Circinus, especially because these two objects both have a disk-like component and a near-polar extended component.
These two moderate-luminosity galaxies are so close that the long baseline interferometric measurements represent a physical resolution that is not obtainable for any of the other survey galaxies.
To include them in our analysis on a comparable basis to the others, we only included interferometric measurements for these two at projected baselines < 40\,m, which only includes information about the extended component and an unresolved component (the disk resolved with higher projected baselines). 
Therefore, it is likely that due to the arbitrary location of the axis system in our analysis, the best-fit model is influenced by the elongations of the sources, and as a result, fixing the opening angle produces the required elongations (1:2) but with an incorrect system axis.

\begin{figure*}
   \centering
   \begin{minipage}{0.5\hsize}
      \centering
      \includegraphics[width=\hsize]{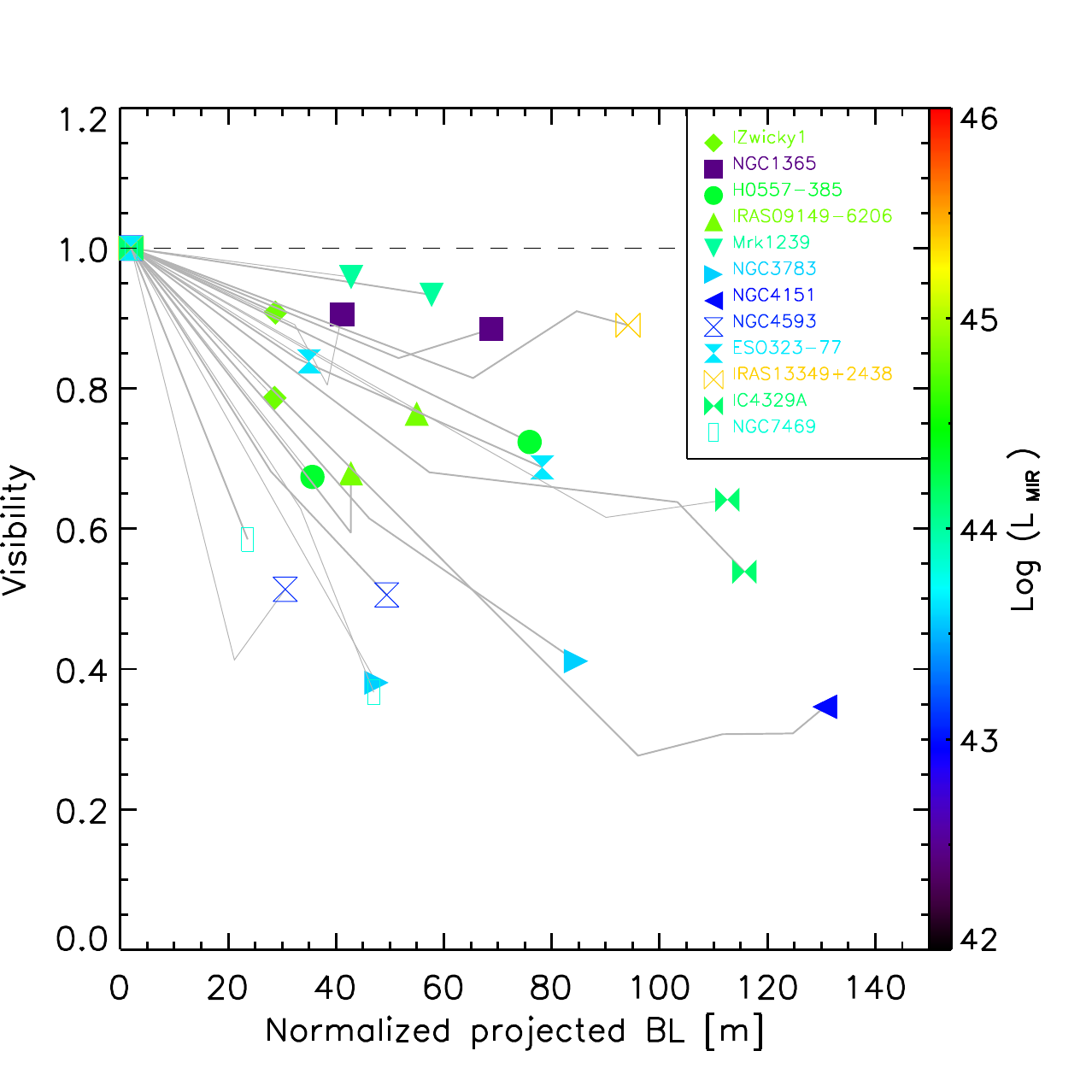}
      \includegraphics[width=\hsize]{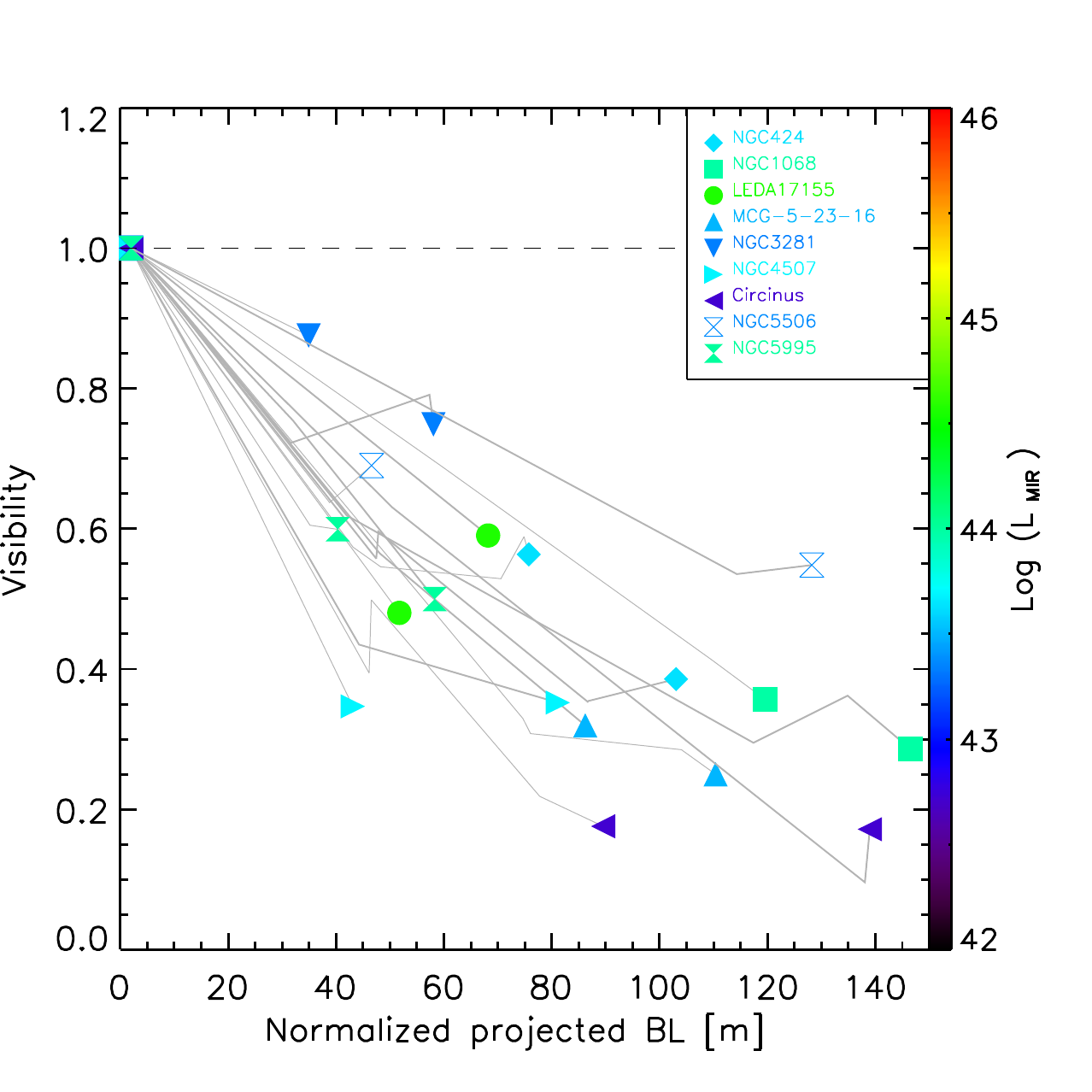}
   \end{minipage}%
   \begin{minipage}{0.5\hsize}
      \centering
      \includegraphics[width=\hsize]{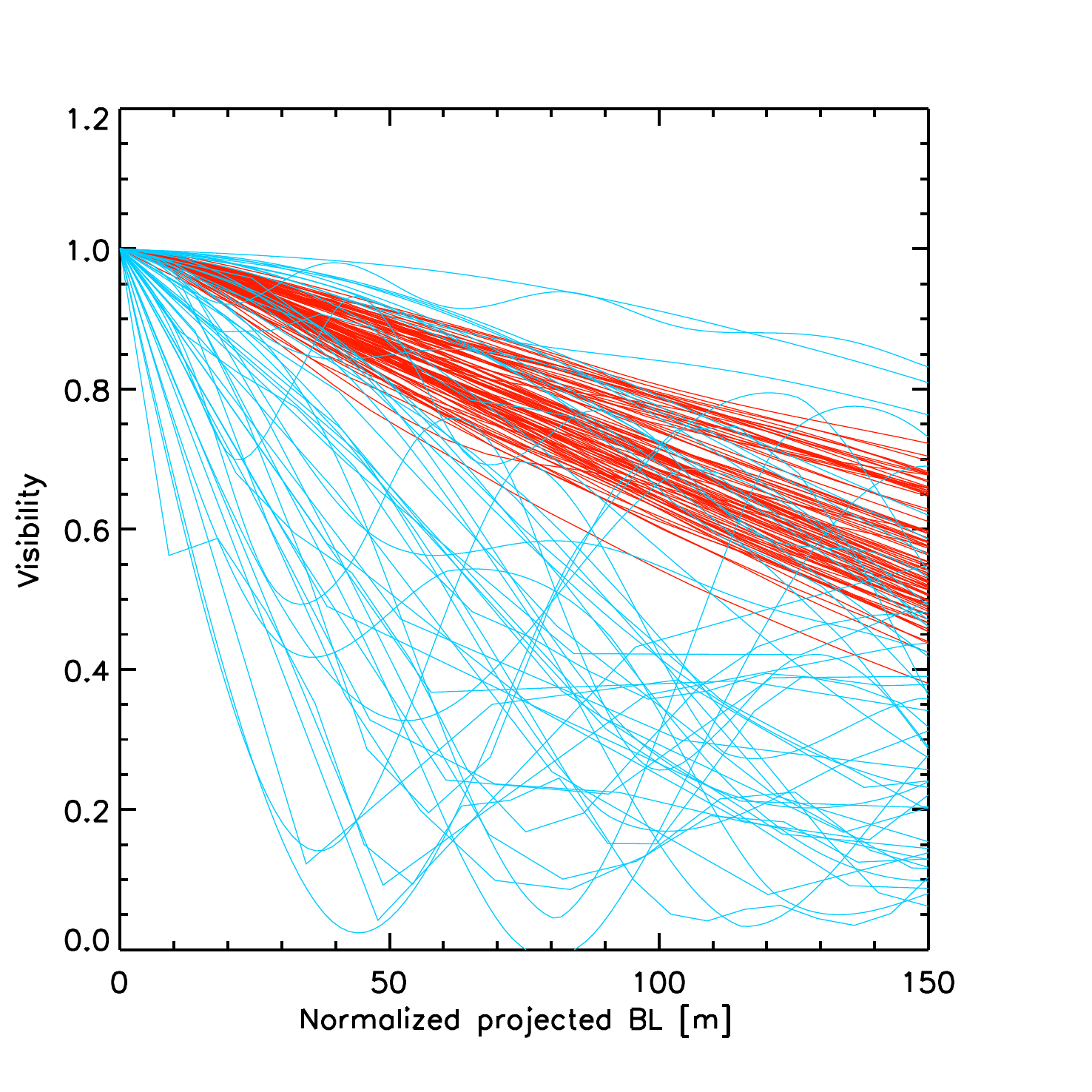}
      \includegraphics[width=\hsize]{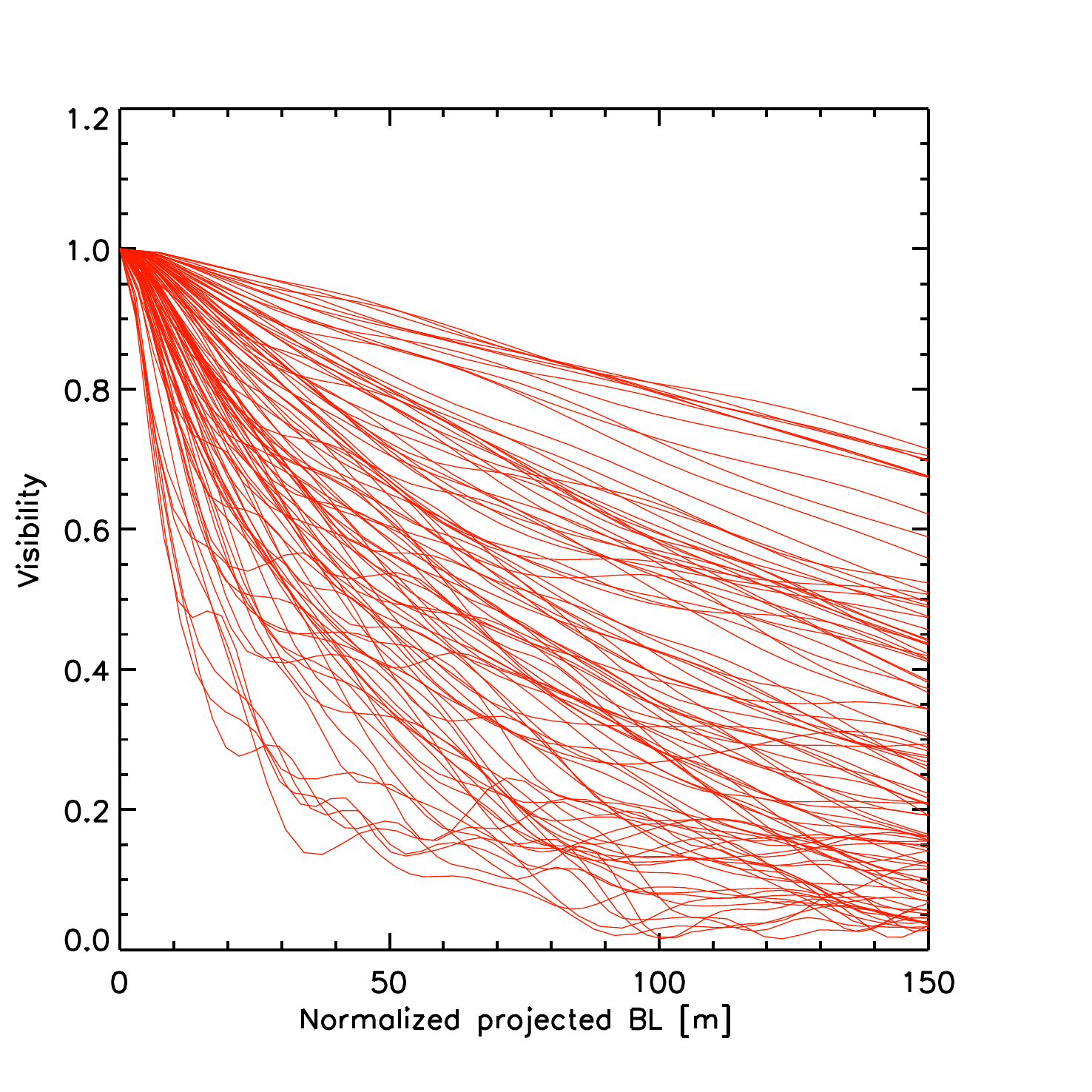}
   \end{minipage}
   \captionof{figure}{{\it \textup{(Left) 12\,$\mu$m interferometric visibilities of type I ({\it top}) and type II sources ({\it bottom}) plotted against the {\it \textup{normalized}} projected baseline.
   For every object we include visibilities for two different position angles connected by independent lines. 
   The normalized baseline is scaled from the observed baseline for each source to normalize its single-aperture 12\,$\mu$m flux; cf. Sect. \ref{subsec:rescaling}.
   Each symbol indicates the longest baseline data point available at the given position angle for an individual object.
   The color of the symbols indicates the value of the infrared luminosity of the source as shown on the scale at the right, data are from by Table~\ref{tab:data}.
   {\it (Top right)} Model normalized 12\,$\mu$m interferometric radial plots for various lines of sight where the nucleus is exposed, corresponding to type I objects, computed from type A models (blue) and from type B models (red). 
   {\it (Bottom right)} Model radial plots for various obscured lines of sight, corresponding to type II objects, computed  for the best type B models.}}}
   \label{fig:rescale}
\end{figure*}

\subsection{Line-of-sight selection}

Since the same input bolometric was used for all the models, using a weight means that  bright images are more likely to be observed, since the ratio between the infrared and UV is lower than faint images in the infrared.
For high optical depths this  causs the images with high self-absorption to be rarer.
We applied the same weighting exponent to both types (I an II) throughout all our work.
Only type II sources show a clear difference when using the reweight.
The 12\,$\mu$m emission of the type I sources is less likely to be affected by self-absorption of the dust. 
The dispersion of the 12\,$\mu$m fluxes for a particular model for type I objects is not very broad, and therefore the reweighting does not play an important role.

For type  II objects this reweighting is quite relevant. For high-inclination values, the 12\,$\mu$m fluxes can be more affected by the self-absorption of the dust clouds. 
In Fig.~\ref{fig:weigamma} we show a comparison for different parameters using a rescaling with $\gamma=1$ and without weights.
The greatest difference is that if we do not use the reweighting, models with low filling factors become  likely for type II sources.
The reason for this is that in these types of models the hot surfaces of the individual clouds produce similar bright spots in the large scales as the surfaces produced by models with high filling factors, where the emission in the large-scale structure is produced by escaping emission through holes.
The rescaling we performed to match the fluxes and distances of the real sources modifies the sizes and  aligns them with those given by our interferometric measurements. Although the geometry generally looks  similar, the problem of not using a reweighting is that for type II models with low filling factors the ratio between the bolometric luminosity and the infrared luminosity becomes extremely high, some of ratios are even quite unrealistic, as seen from the luminosity function of Seyfert galaxies.

\section{Discussion}
\label{sec:discussion}

\subsection{What does the interferometer see?}
\label{subsec:what-do-we-see}

  \begin{figure*}
\centering
\begin{minipage}{\textwidth}
  \centering
   \includegraphics[width=\hsize]{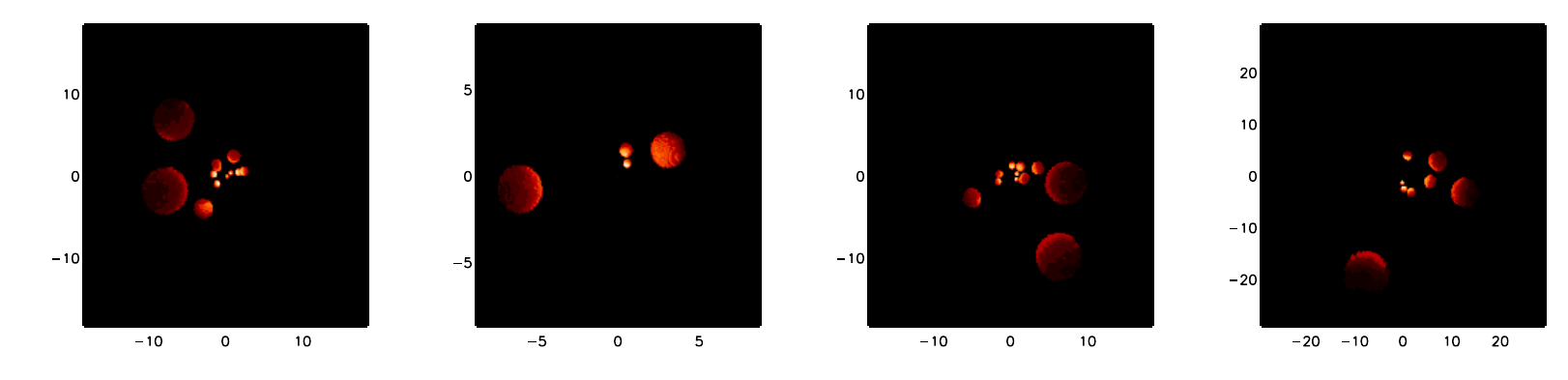}
     \captionof{figure}{{ 12\,$\mu$m images created from one of the best type I models (model A). We only show lines of sight where 
     the nucleus is exposed. Each plot shows a different realization of the cloud positions.
     Labels denote the distance to the center in pc.}}
  \label{fig:imaget1_A}
  
   \includegraphics[width=\hsize]{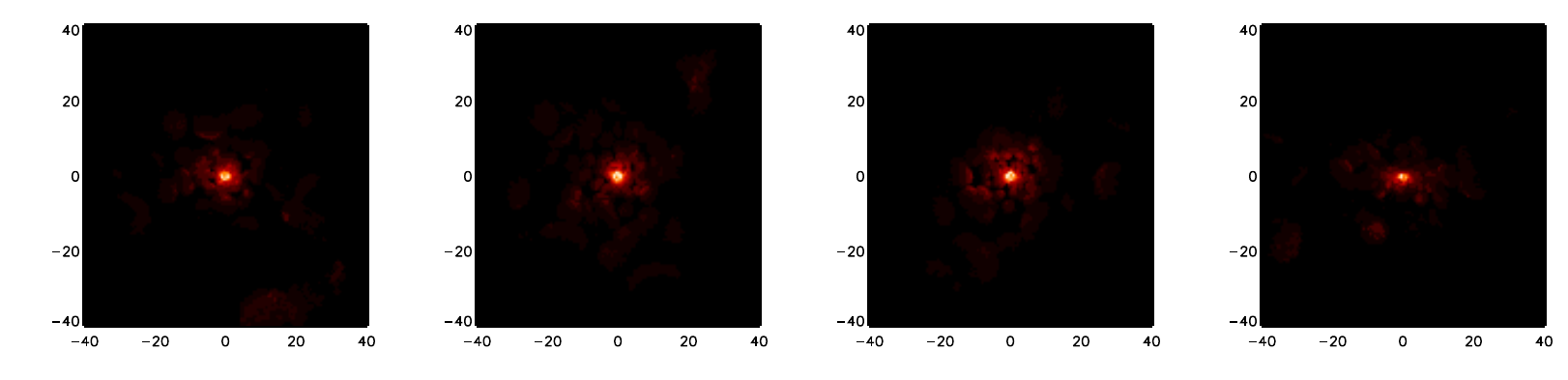}
     \captionof{figure}{Same as Fig.~\ref{fig:imaget1_A}, but the images are created from the unobscured lines of 
     sight from model B, i.e., type Is with the same filling factor as type II objects.}
  \label{fig:imaget1_B}

   \includegraphics[width=\hsize]{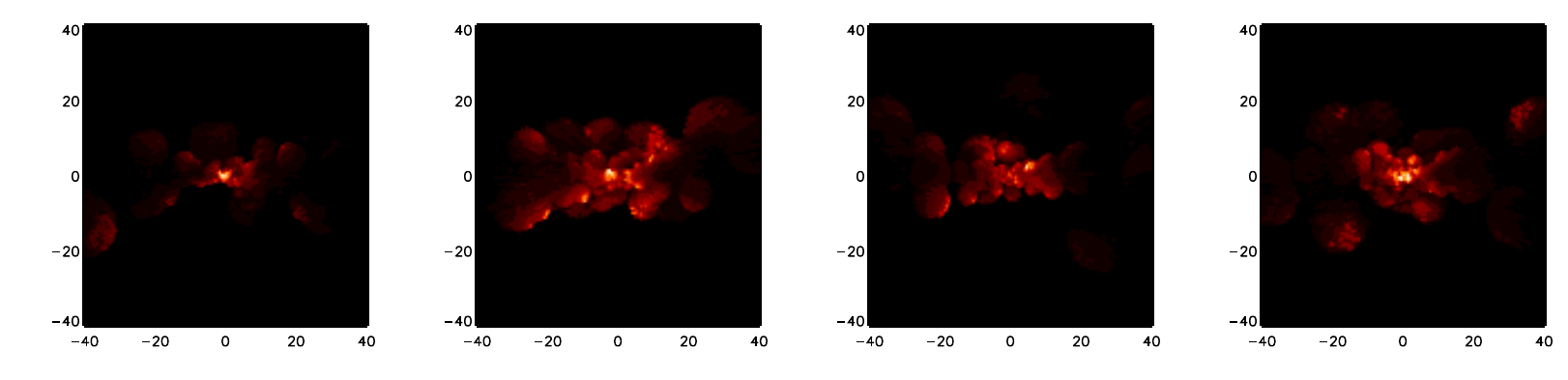}
  \captionof{figure}{Same as Fig.~\ref{fig:imaget1_A}, but the images are created from the obscured lines of 
     sight from model B, i.e., the best type II model.}
  \label{fig:imaget2_B}
\end{minipage}%

\end{figure*}
 
In this section we examine the images of the best models to acquire intuitive insight into their structures, and to understand which features in the models cause noticeable differences in the actual observations.


Our work shows that apparent differences in the mid-infrared morphology arise not only from inclination effects, but that statistical variations in the cloud distribution can be relevant as well.
When the size of the clouds is large enough and the fraction of the volume occupied by the clouds is relatively low, the appearance of the mid-infrared emission will vary depending on our specific line of sight and realization of the models \citep{2006A&A...452..459H, 2008A&A...482...67S}. 
In the probabilistic models presented by \citet{2008ApJ...685..147N, 2008ApJ...685..160N},  these variations do not appear explicitly because their models are built using average quantities, therefore differences that are due to statistical variations of the clouds are ignored. 

In Fig.~\ref{fig:rescale} (top) we plot the observed interferometric 12\,$\mu$m visibilities of our objects using measurements along two distinct position angles if available. The baselines are rescaled to compensate for their luminosities and distances, as described in Sect.~\ref{subsec:rescaling}.
To the right of the same figure we plot the model visibilities for different realizations {of type A and type B models} that would have been classified as a Seyfert I galaxy because the nucleus is directly visible.
In Fig.~\ref{fig:imaget1_A} we show model images of four realizations of type A models and in Fig.~\ref{fig:imaget1_B} images of type B models with unobscured nuclei.
The lower plots in Fig.~\ref{fig:rescale} and the images in Fig.~\ref{fig:imaget2_B} represent observations and models of Seyfert II galaxies and type B models with obscured nuclei.
Except for two objects in the left top plot of Fig.~\ref{fig:rescale}, when normalized in the infrared, low-luminosity objects seem to be better resolved than high-luminosity objects.

The plot of Fig.~\ref{fig:rescale} (top left) shows large variation in visibilities of the Seyfert I galaxies, which is reproduced by the low filling factor type A models.
The type B unobscured models show much less variation and relatively high visibilities because more clouds in the model are located closer to the inner regions of the torus, making is seem compact and smooth.
Figure~\ref{fig:imaget1_A} shows that the appearance of the low filling factor models is determined by the positions of a few hot, bright, unobscured clouds around the nucleus.
The random variations of the positions of these clouds in the realizations creates the large variations in apparent visibility.
This creates the apparently uniform high visibilities in these models.
It is clear from the plots that the curves from the high filling factor type B model cannot reproduce the overall distribution of observed visibilities for our full sample of type I objects: the variation in visibilities would be too low.
Objects such as NGC~3783, IC4239A, and NGC~4593 have large dispersions in their visibilities that cannot be explained with the type B model. 
Although the aim of this work is not to find the physical explanations for the dusty structure, we note that the three mentioned objects have lower luminositis than the less resolved objects (e.g., IRAS~13349+2438). 
We cannot, of course, exclude that some of the type I galaxies represent unobscured type B geometries, a situation similar to the original Standard Model.

For Seyfert type II galaxies, images from obscured lines of sight of type B models are shown in Fig.~\ref{fig:imaget2_B}.
With the nuclear regions blocked by dust, the emission is dominated by the accidental positions of relatively free lines of sight through holes in the cool dust to warmer areas at various radii.
These accidents produce the variations in visibility seen in the bottom plots of Fig.~\ref{fig:rescale}.
Once again, a few of the observed Seyfert II galaxies may arise from type A low-density geometries where the line of sight is blocked by a stray cloud.

\subsection{Spectral energy distribution.}
\label{subsec:SEDs}

We only analyzed the N-band data, where the new interferometric measurements include more spatial information than the single-aperture spectral energy distributions (SEDs) alone. 
Ideally, we should describe the SED and interferometric data simultaneously. 
We did not attempt this because of the difficulties of consistently calibrating multiwavelength observations, the very different resolutions and fields of view of these observations,  possible contamination from other physical sources, and lack of multiwavelength observations for most of our objects. 

In spite of these problems, we can produce the SEDs with the radiative transfer code over broad wavelength ranges for our best-fit models with the purpose of displaying the overall predicted behavior of the spectra. 
The SEDs presented in this work can be further investigated by us or other groups to verify them outside the MIR window. 
In particular in the near-infrared, the promising technique presented by \citet{2015A&A...578A..47B} for isolating the NIR emission of the AGN is expected to provide good constraints for the hot emission.

\begin{figure}
   \centering
   \begin{subfigure}{\hsize}
      \includegraphics[width=\hsize]{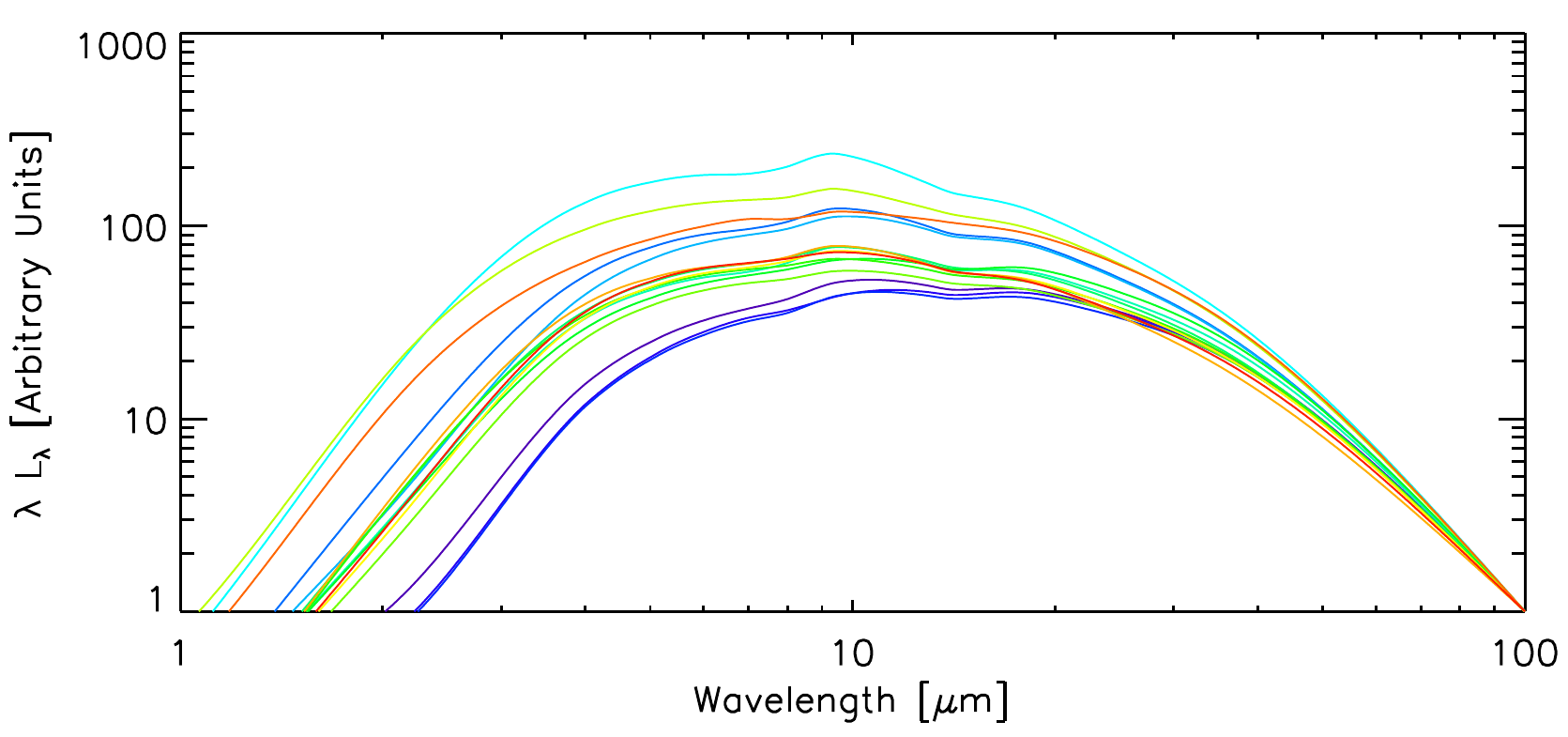}
      \caption{Using the best-fit model of type I objects. 
      Only unobscured lines of sight are considered here.
      The contribution of the accretion disk is not included in the SED.}
      \label{fig:sedty1}
   \end{subfigure}

   \begin{subfigure}{\hsize}
      \includegraphics[width=\hsize]{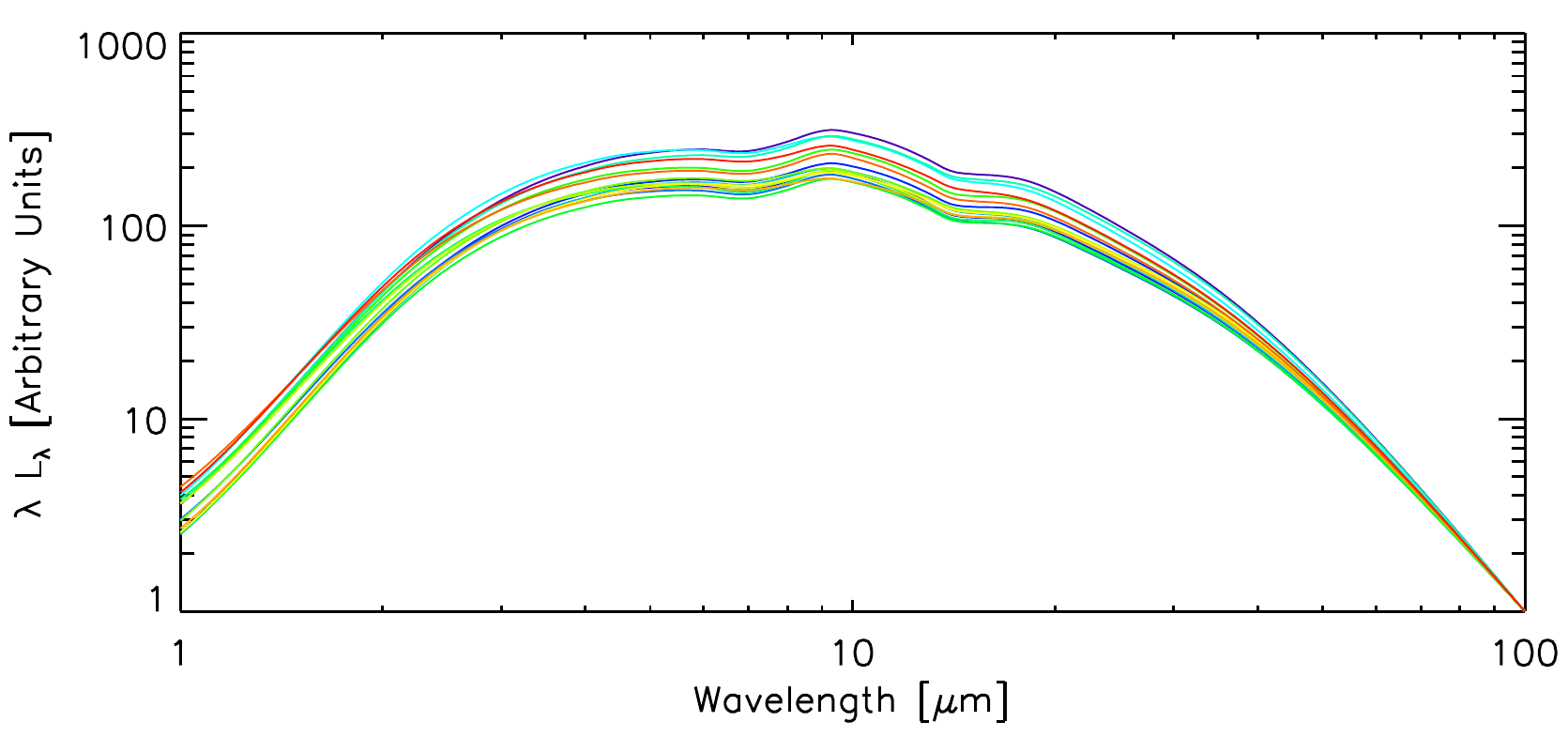}
      \caption{The same as Fig.~\ref{fig:sedty1}, but for unobscured lines of sight from the type II model.}
      \label{fig:sedty2_1}
   \end{subfigure}

   \begin{subfigure}{\hsize}
      \includegraphics[width=\hsize]{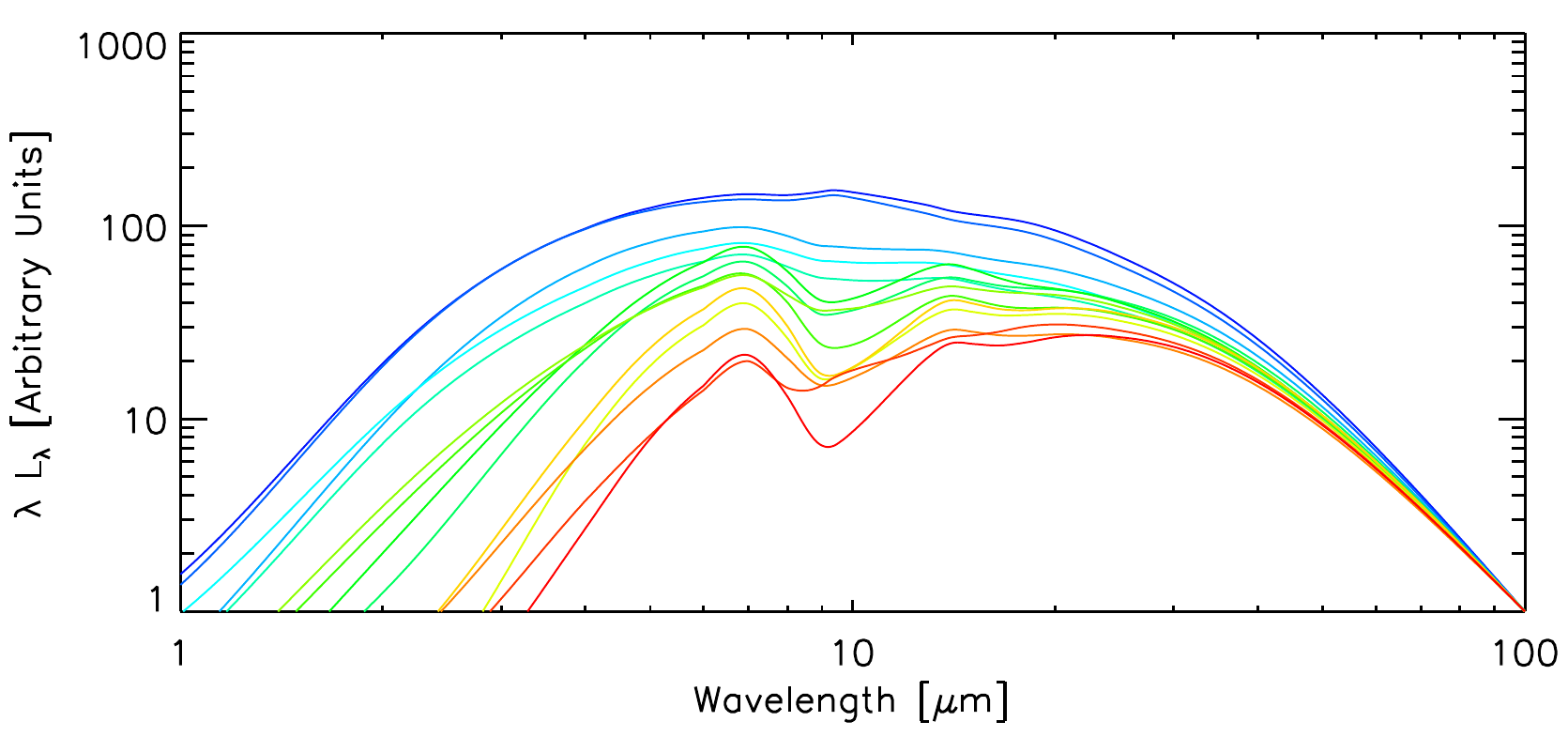}
      \caption{Same as Fig.~\ref{fig:sedty2_1} for obscured lines of sight from type II models.}
      \label{fig:sedty2_2}
   \end{subfigure}
   \caption{SEDs for multiple realizations of the best-fit model. }
\end{figure}

We show examples of different realizations of the best type I model for unobscured inclinations (in Fig.~\ref{fig:sedty1}) and for obscured inclinations using the best type II model (Fig.~\ref{fig:sedty2_2}). 
We also include the SEDs of unobscured inclinations for the best type II model in Fig.~\ref{fig:sedty2_1}.
The SEDs corresponding to different realizations of obscured type II objects show a diverse family of spectra with variations of the silicate feature in absorption. 
Similar to other torus models, the modeled spectra only show a moderate absorption feature in contrast to the deep silicate feature typically present in continuous models.
From Fig.~\ref{fig:sedty2_2} we observe that it is also possible to obtain SEDs with relatively high emission at short wavelengths from hot dust coupled with small silicate absorption features. 
It is quite likely that such SEDs correspond to regions with holes in the cloud distribution through which the hot emission from the inner regions is seen, giving rise to a significant contribution of flux in the near-infrared. 
The small silicate feature in this case could be explained as an average between the absorption feature produced from the back faces of the clouds and the silicate feature in emission that is viewed through the holes of the torus. 
This could explain the absence of a silicate feature in absorption and the relatively blue spectra of NGC~424 described by \citet{2012ApJ...755..149H}.

The main differences between the SEDs of the true type I models and those of the unobscured type II models are in the strength of the silicate feature and the slope of the spectrum in the 2\,--\,20\,$\mu$m wavelength range.
The silicate features in true type I models vary from weak absorption to moderate emission; the spectral slopes vary from moderately hot (large near-IR contribution) to moderately cool; the warmth of the continuum slope is directly correlated with the strength of the emission feature.
For the unobscured type II models the silicate feature is always seen weakly in emission and the continuum spectrum (in units of $\lambda L_\lambda$) weakly rising toward shorter wavelengths.

\begin{figure*}
   \centering
   \begin{subfigure}{0.32\hsize}
      \includegraphics[width=0.98\hsize]{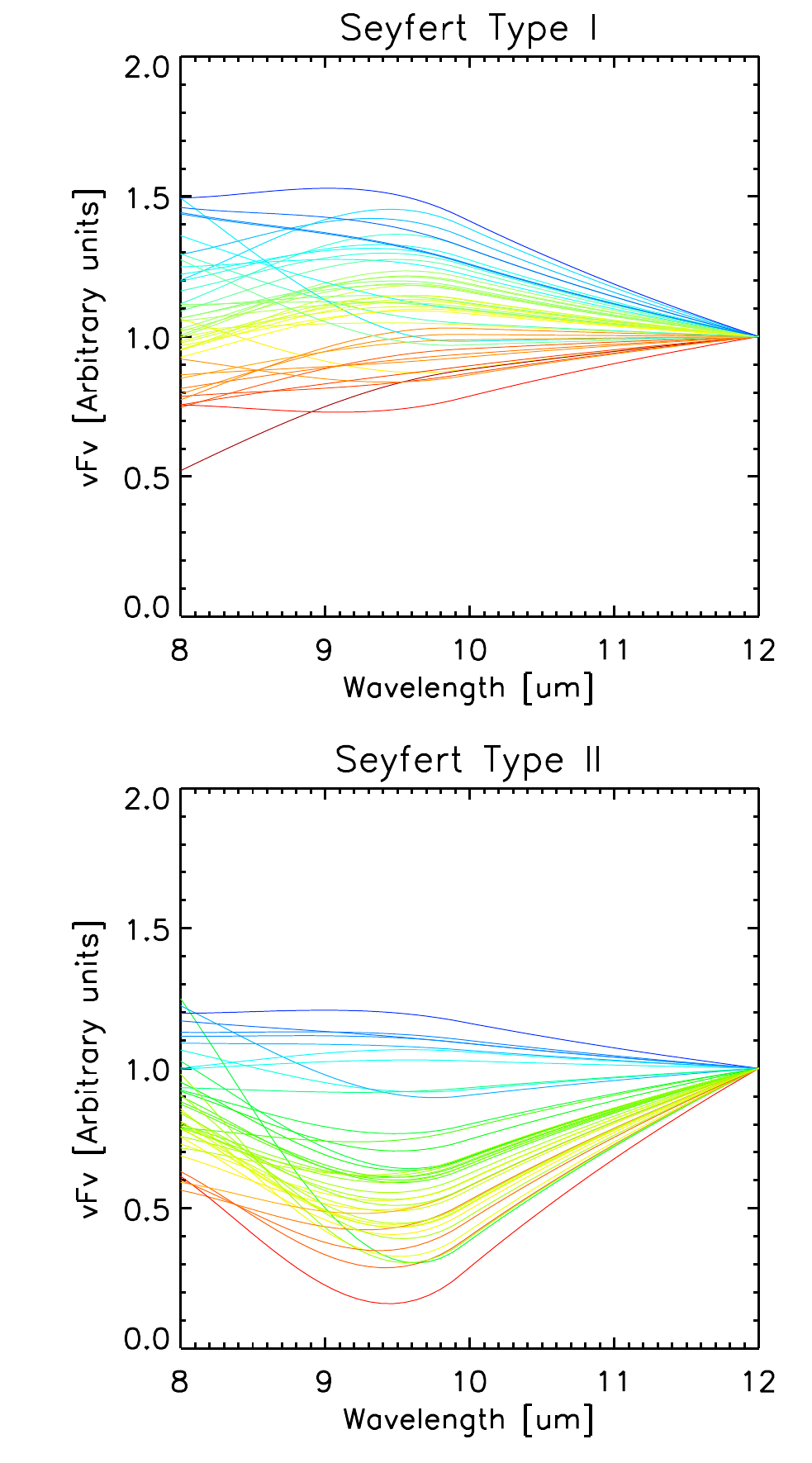}
      \caption{Spectra from best-fit models.}
      \label{fig:NbandModels}
   \end{subfigure}
   \begin{subfigure}{0.32\hsize}
      \includegraphics[width=0.98\hsize]{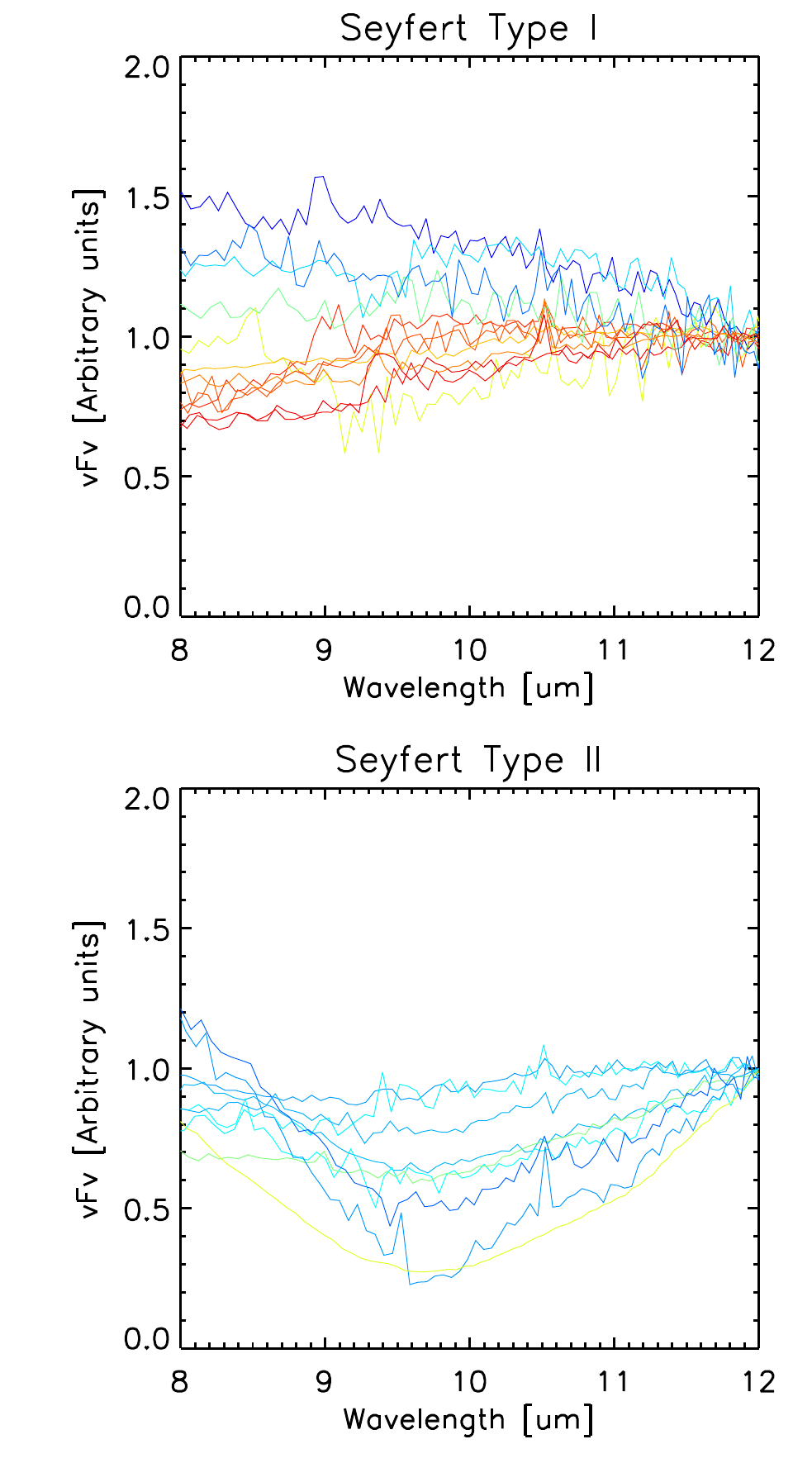}
      \caption{Observed objects: Single-aperture spectra.}
      \label{fig:NbandSingle}
   \end{subfigure}
   \begin{subfigure}{0.32\hsize}
      \includegraphics[width=0.98\hsize]{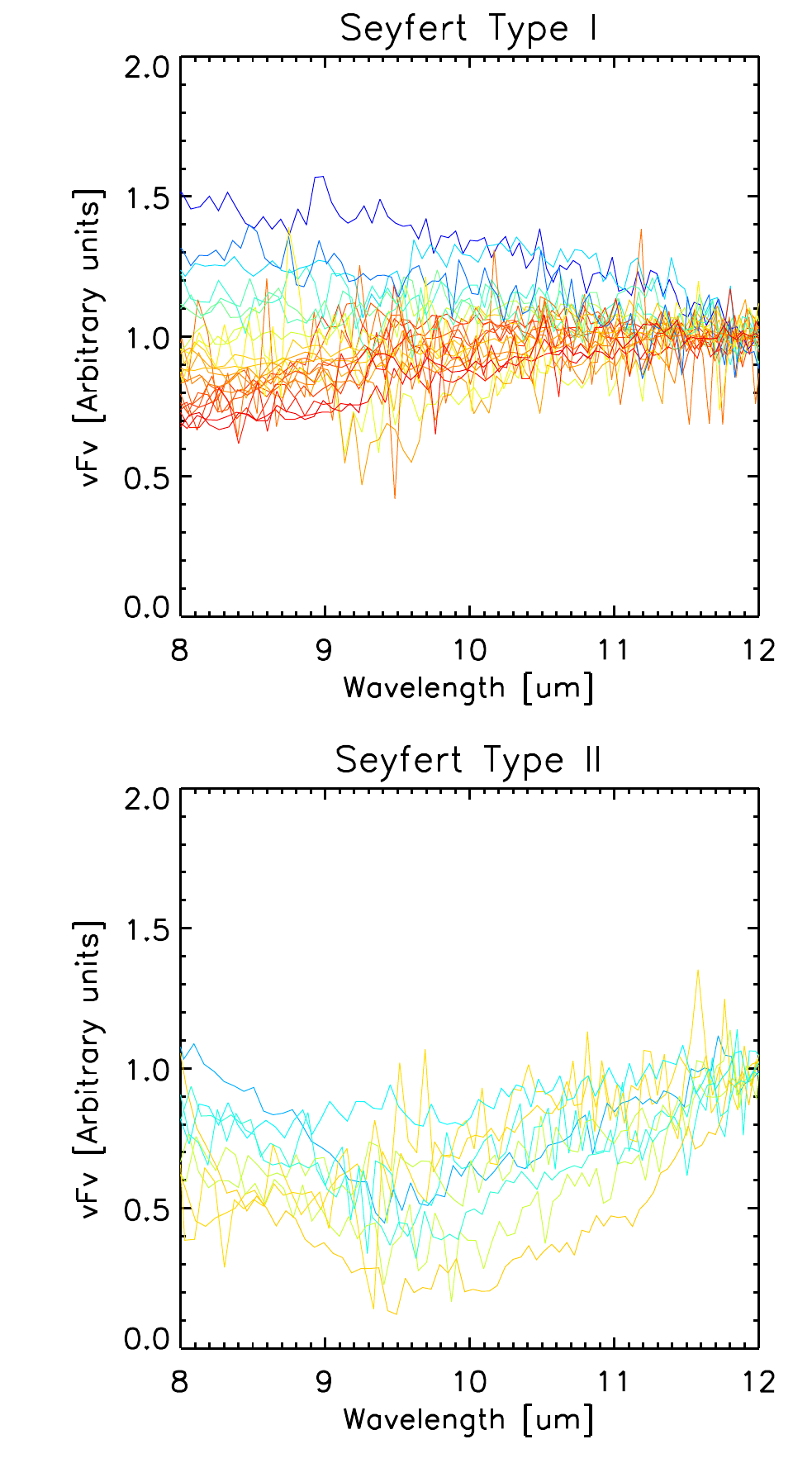}
      \caption{Observed objects: lowest baseline MIDI spectra}
      \label{fig:NbandMidi}
   \end{subfigure}
      \caption{N-band spectra for type I objects ({\it top row}) and type II objects ({\it bottom row}). 
      The spectra have been normalized to the 12\,$\mu$m flux. 
      The colors indicate the slope of the continuum. }
\end{figure*}

In Fig. \ref{fig:NbandSingle} we zoom into the 8\,--\,12\,$\mu$m single-aperture spectra for the observed objects, as well as the output spectra from their respective best-fit models.
The multiple spectra are generated from  different realizations and multiple inclinations of the best-fit model.
In the top row we observe that the spectra of type I objects agree well with the predictions of the model. 
The diversity of slopes and the featureless spectra seem to be well described.
As a comparison, we additionally include the interferometric spectra from the lowest baseline available. 
Many of our type I objects are slightly resolved with the shortest baseline resolution, so the differences in the shape of the spectra are small.
If contamination by surrounding starburst regions is present in the single-aperture spectra, however, the shortest baseline spectrum should be less affected by this.
The observed type II spectra are  also well reproduced by the best-fit modeled spectra. 
Objects with deep silicate features can be explained with our model, although they are less common.

\subsection{Are type Is different from type IIs?}
\label{subsec:different}

The strictest form of the AGN Standard Model explains all differences between the Seyfert types by line-of-sight effects. 
This model assumes that the dusty tori of all Seyfert galaxies have very similar properties.
Our attempts to model the mid-infrared interferometric data indicate that this is not possible. 
To fit the observed sample we need (at least) two different models, distinguished
primarily by different dust-filling factors in the volume radiating in the mid-infrared.
For this subsection only we denote for brevity the low filling factor models, consistent with the type I galaxies, as type A models and the high filling factor models as type B.

Considering all { lines of sight (LOS)}, approximately 10-30\,\% of the type A models  would be classified as Seyfert II galaxies by optical observers because the LOS happens to hit a cloud.  
Conversely, for the best-fitting type B model, approximately 40\,--\,50\,\% would be classified as Seyfert I because the LOS allow a direct view of the nucleus, either because it lies within the torus opening angle or by chance misses all clouds.
We also note that although the type I and type II source subsamples  as a whole require different models, there are individual sources that can be described with either model.  
These considerations bring back the Standard Model in a weakened form.  
While most of the observed Seyfert II galaxies have model B structures, some of them have model A structures, but are classified as Seyfert II because of the viewing geometry, and
vice versa for Seyfert I galaxies.  

Our result of the intrinsic differences between type I and type II sources in terms of the filling factor or covering fractions was previously suggested by the results of \citet{2011ApJ...731...92R}, who used fits on the SEDs of individual galaxies. 
With our method, we proceed from the usual SEDs studies by using high-resolution data provided by interferometers where the emission is indeed being resolved and by finding models that statistically reproduce the general features of a sample of sources instead of focusing on the details of individual objects.
With our results, we support the statement that the covering fraction of the torus should be lower for type Is than for type IIs (\citet{2012ApJ...747L..33E}. 

We do not go in detail into the question of the true percentages of high or low filling factor structures in the local Universe because this requires extremely careful consideration of how any observational sample is selected, but we discuss some of the consequences of accepting two different underlying structures.
We assume that half of the Seyfert galaxies observed are type II.
If half of the type B structures are classified as type I sources (e.g., because they
are observed within the opening angle), then the fraction of intrinsic type A structures must be small, otherwise the fraction of galaxies classified as type I would exceed 50\,\%.
But this situation is essentially that of the Standard Model and contradicts our main result that the type B models fail to fit globally the interferometrically observed sample of type I objects.

If we reduce the fraction of type B structures that are classified as Seyfert Is to below 50\,\%, then the fraction of true type As among the Seyfert Is of course rises.  
If we require that $>$50\,\% of the Seyfert Is be in fact type A structures (to be consistent with our interferometric measures), the maximum fraction of type B structures that cross over in the observations is 30\,--\,40\,\%.

\subsection{Mid-infrared emission efficiency}
\label{subsec:ir-xray-rel}

In this section we try to find a reasonable estimate of the intrinsic amount of UV flux, emitted by the accretion disk, by using the observed 12\,$\mu$m nuclear flux and the efficiency ratio $\eta_{UV-IR}$, defined in this case as the ratio between the UV flux and the observed 12\,$\mu$m flux.
For a dusty medium distributed non-uniformly in a volume, the efficiency ratio is no longer constant, but is dependent on the line of sight.
The variations of the efficiency ratio $\eta_{UV-IR}$ depend on the distribution of the medium and in particular become larger when the line of sight is optically thick in the infrared, or in other words, when self-absorption of the dust becomes more relevant. 
The diversity of $\eta_{UV-IR}$ values should be kept in consideration when computing the UV luminosity from the observed infrared luminosity in Seyfert galaxies. 

\begin{figure}
   \centering
   \includegraphics[width=\hsize]{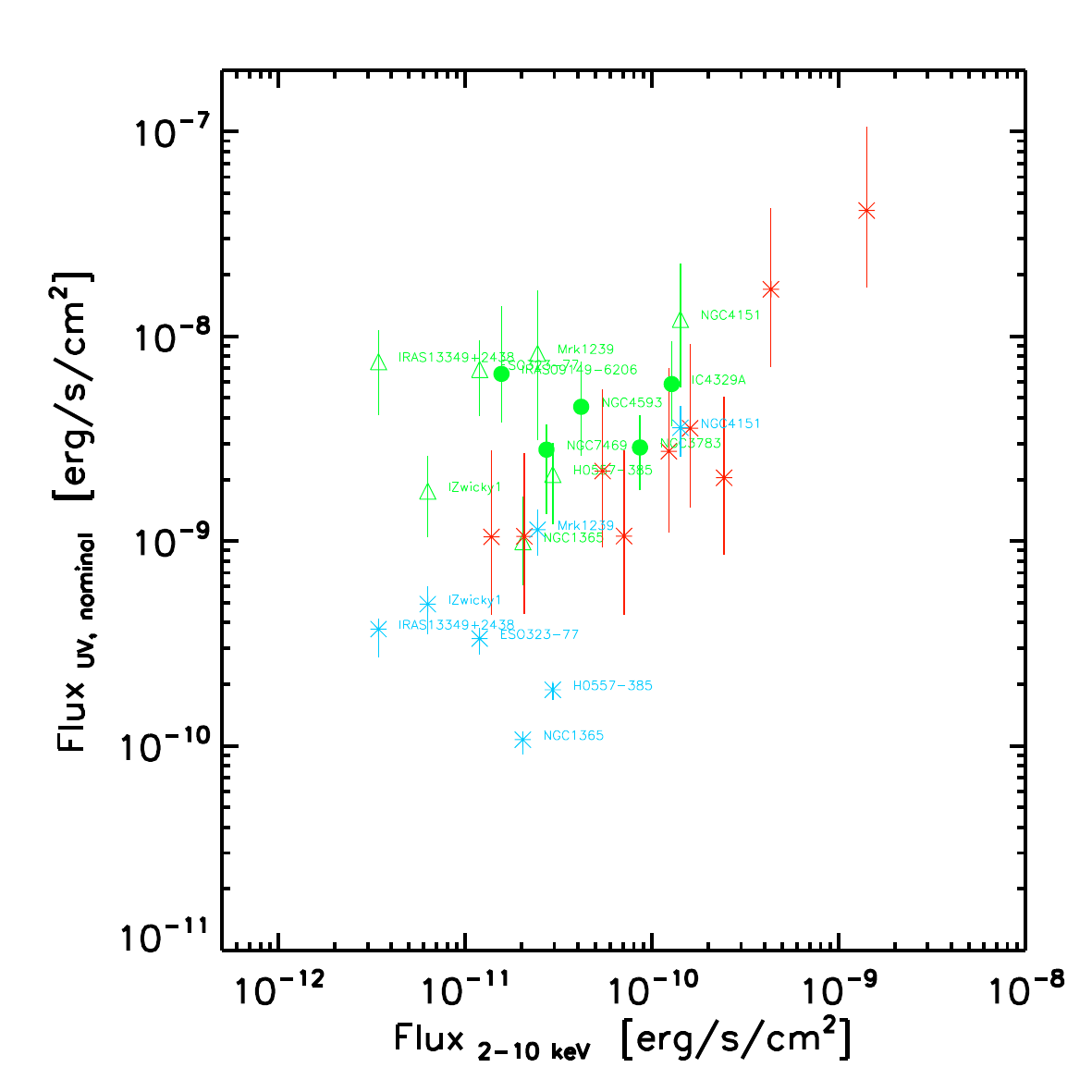}
   \caption{Absorption-corrected  $2\,--\,10$\,keV X-ray flux versus nuclear UV flux estimated from models. 
   The symbols indicate the median value for the estimated UV luminosities from their respective best-fit models, and the lines indicate the dispersion (68\,\% area) in the possible values. 
   Type I galaxies that can only be fit with type A models (filling factors 0.4\,--\,1.4\,\%) are given as filled circles. 
   Those that can also be fit by type B models (filling factors 5.3\,--\,20\,\%) are shown twice, with the model UV-flux values given as triangles (model A) and asterisks (model B).
   All type II galaxies are shown as asterisks.}
   \label{fig:xrayuv}
\end{figure}

To obtain a reasonable estimate of UV flux for our objects, we computed for every best-fit model the distribution of the efficiency factors along multiple line of sights and with different realizations. 
For the type II objects we computed the distribution of the efficiency ratio for model B (best type II model) reported in Table~\ref{tab:parameters}, and then we used the infrared flux to obtained an estimate of the UV emission from the accretion disk. 
We show the computed UV flux in Fig.~\ref{fig:xrayuv} together with the corrected 2\,--\,10\,keV X-ray fluxes.

For the type I models we used a slightly different approach.
We previously showed that our entire sample of type I objects cannot be fit with the same model as the type IIs, but it might be  possible that a fraction of our type Is can be consistent with the unobscured lines of sight of the type II model.
For objects where our model B fails in describing the interferometric measurements we seem to find a good fit using a low filling factor environment. 
For our type I objects we individually searched for the best-fit models using a low filling factor between 0.4\,--\,1.4\,\% and also a higher filling factor between 5.3\,--\,20\,\%.
In Fig.~\ref{fig:xrayuv} we show for every type I object two estimates of the UV flux, one using a model with a low filling factor and the second using a high filling factor.
For objects that cannot reasonably be described using a high filling factor model, we only show the estimates from the low filling factor model. 

Corrected 2\,--\,10\,keV X-ray luminosities are assumed to be related to the UV bump of the accretion disk and are sometimes used to estimate the UV luminosity.
From Fig.~\ref{fig:xrayuv} we observe that a correlation might exist if we allow a mixture of objects with a low and a high filling factor.
It is also clear that our  objects with the low filling factor models do not seem to be outliers despite their general low efficiency ratio $\eta_{UV-IR}$.

\subsection{Stability of the clouds}

Our best-fit low filling factor models  are built with a limited number of clouds with high optical depths, typically with $\tau_{9.7} \geq 8$. Clouds with such optical depths should be quite massive, therefore the question arises whether these clouds are physically possible. 
In particular, we investigate if the thermal pressure is sufficient to keep the clouds from collapsing. 
To answer this question, we  used the Jeans instability criterion. 
From the virial theorem and assuming a static, spherical, homogeneous cloud, we obtain that the critical mass for a cloud to collapse is given by $M_J \sim 6.64 \times10^{22} T^{3/2} \rho^{-1/2} [g]$. 
If the mass of our cloud exceeds the value of $M_J$, the thermal pressure is not enough to keep our cloud from collapsing. 

Since for our best type I model the clouds are located close to the inner rim, we tested the stability of one such clouds. 
The typical volume of a cloud close to the inner rim is $V_{cl}=9.8\times 10^{53}$\,cm$^3$ and the dust mass of the cloud is  $M_{dust}=0.77$\,M$_\sun$.
To derive the total mass of the cloud we used a dust to gas ratio $\rho_{dust}/\rho_{gas}=0.01$, which gives a total mass of $M_{tot}=77$\,M$_\sun$.
From the radiative transfer computation we obtain an average temperature of the cloud of $T\sim 270$\,K. The resulting Jeans mass for this configuration is $M_J\sim 361$\,M$_\sun$, which is greater than the total mass of our cloud $M_{tot}=77$\,M$_\sun$. 
Although this calculation does not include other effects, such as rotation or shearing effects, we show that in a static situation the thermal pressure can prevent such clouds from collapsing. 
In fact, our calculation predicts that the cloud might expand since it is not confined by gravity. 
It might also be possible that the clouds are confined by external gas pressure. 

\section{Conclusions}
\label{sec:conclusions}

We presented results from a statistical method developed to interpret interferometric
data with complex radiative transfer models.  
We applied our method to the interferometric data of AGNs published by \citet{2013A&A...558A.149B}
and constructed our model images according to the dusty torus models from \citet{2008A&A...482...67S}.
We summarize our major findings below.

\begin{enumerate}
   \item Mid-infrared interferometric data of a combined AGN sample, including both type I and type II sources, cannot be described by a single stochastic model (using \citet{2008A&A...482...67S} models) under the assumptions of the Standard Model where observed differences are only attributed to inclination and line-of-sight effects. 

   \item Type I and type II sources can be well explained by such models if they are taken as two separate subsets with different model parameters for each subset. 
   We found that the greatest difference between the models that describe each subset is  in the volume fraction that the clouds occupy in the inner regions. 
 
   \item Seyfert type I galaxies are best explained by using torus models with low filling factors at the inner regions, between 0.4\,\% and 1.5\,\% of the volume of a spherical shell.
   The low filling factor implies a relatively small number of clouds.
   This small number produces large apparent fluctuations in interferometric measures of the type I sources, including a broad range of apparent geometrical sizes.
   This agrees with the large dispersion in sizes reported by Burtscher et al. (2013).
   
   \item Seyfert type II galaxies are best explained with torus models with a filling factor  of 5 or larger than those describing the Seyfert type Is.
   The torus emission in the type II sources seems to be dominated by the warm infrared emission from a very compact region that escapes through the holes created by the clumpy nature of the torus.
   These random holes might be causing the asymmetrical emission in the large-scale structure.

   \item Although two models are necessary and sufficient to explain our observations of  the two Seyfert subsets, this represents an oversimplification.  
   By accidents of obscuration, some of the observed type I sources may arise from high filling-factor geometries and some of the type II sources from low filling-factor geometries in a more complicated version of the Standard Model.
   In addition, of course, more than two geometries may actually be present.
 
   \item The reduction of the silicate feature in our models is mostly caused by the large optical depth of the clouds and to a lesser degree to the shielding effect caused by non-silicate grains.
   For a low number of clouds, the reduction  of the silicate feature is not caused by outer clouds blocking our view to the hot surfaces of the inner clouds.

\end{enumerate}

\begin{acknowledgements}
The authors thank the referee  for the thoughtful and helpful comments. 
We also thank S. Honig, K. Meisenheimer, M. Schartmann, and L. Burtscher for their comments, discussions and help, which all contributed to making this work possible.
N. L\'opez-Gonzaga was supported by grant 614.000 from the Nederlandse Organisatie voor Wetenschappelijk Onderzoek and acknowledges support from a CONACyT graduate fellowship.
\end{acknowledgements}

\bibliographystyle{aa} 
\bibliography{/home/nlopez/Documents/PhD_thesis/Template_Silv/BackMatter/Bibliography/bib/Bib_silvia}

\newpage
\begin{appendix}

\section{Model setup}

The models used for our database of infrared images are built based on the approach described by \citep{2008A&A...482...67S}. 
These  wedge-like clumpy torus models are one of many different torus models currently available, but their main advantage is that it is relatively easy to proceed from a model with only a few clouds to models with a large number of clouds that resemble the smooth distribution of continuous models better. 

The models are built using spherical coordinates. 
The dust-free volume is defined by the half-opening angle $\theta_{op}$, where clouds are only allowed to exist within the region of $\theta_{op} < \theta < \pi - \theta_{op}$. The cloud centers are distributed in equal volumes randomly along the azimuthal direction and polar angle in the allowed zone. 
The radial position of the clouds are randomly distributed and follow a power-law density profile $\rho_r = \rho_0 (r / 1$\,pc$)^{\alpha}$, where $\alpha$ is the density profile index and $\rho_0$ a normalization constant.

Dust clouds are spherical, homogeneously filled with dust, and all possess the same optical depth. 
The radius of the clouds is proportional to their radial position $a_{cl}= a_0  (r/ 1$\,pc$)$, where $a_0$ is a constant value. 
The number of clouds of the model is determined by the filling factor. 
We define the filling factor as the ratio between the volume occupied by the clouds and the total volume of a spherical shell defined by the inner radius of the model and the radius at 1\,pc.  
Finally, the total amount of dust in the model is determined by normalizing the total density in order to obtain a fixed average optical depth at 9.7\,$\mu$m along the equatorial plane.

Since the true mixture of grains in AGNs is not fully determined, we use a typical mixture of dust grains for the intrastellar medium, consisting of 62.5\,\%  silicates and  37.5\,\% of graphites, where the percentages correspond to the mass fraction. 
In the case of the graphites, we take two different sets of optical constants: one third of our graphites is represented by graphites whose electric field vector oscillates in parallel to the crystal axis of the grain, and two thirds of the grains have a perpendicular oscillation.
For the size distribution we use the classic MRN-model \citep{1977ApJ...217..425M}. 
Following \citep{2008A&A...482...67S}, we use a decoupled computation of the temperature for each dust species and grain size. For each dust species we take five bins of different sizes, so we take in total 15 different dust density grids as input.
Since the treatment of the dust temperature is decoupled for each grain size, we also implement the sublimation temperature of each grain type. 
We take a sublimation temperature of 1500\,K for the graphites and 1000\,K for the silicates. 

To approximate the SED of the accretion disk, we use a broken power-law spectrum as described by \citet{2006A&A...452..459H}, which is derived from quasi-stellar object spectra \citep{1998A&A...331...52M}:
\begin{equation}
  \lambda F_\lambda \approx \begin{cases}
    \lambda & \lambda< 0.03\,\mu\text{m}\\
    \text{constant} & 0.03\,\mu\text{m} \leq\lambda \leq 0.3\,\mu\text{m}\\
    \lambda^{-3} & 0.3\,\mu\text{m} <\lambda.
  \end{cases}
\end{equation}

\section{Scaling of the observables}
\label{sec:scaling}

Our procedure for stochastically simulating a specific observation according to a specific set of model parameters is the following: 
\begin{enumerate}
   \item Choose a random cloud realization in accordance with the model parameters.
   Choose also an inclination angle $\theta$ and rotation angle $\phi$ on the sky.
   
   \item Given the nominal model luminosity $L_m$, compute the cloud temperature distribution and three-dimensional radiation field.
   
   \item Using $\theta$ and $\phi,$ project the emitted radiation at the three chosen wavelengths onto a plane with these inclination and rotation angles at the nominal distance $D_m$.  
   Take the two-dimensional Fourier transform of these images to evaluate the model-correlated fluxes at all baselines $BL_m$.
   Evaluate especially the total zero-baseline 12\,$\mu$m flux density $f_m(12)$ and also determine from the optical depth in the visual $\tau_V$ whether this realization would be classified as type I or II.
   
   \item Now consider each actually observed galaxy in the sample, with its actual observed values of $D_s$ and $f_s(12)$.  
   If it is the incorrect Seyfert type, skip this realization.
   Otherwise:

   \item Move the model from $D_m$ to $D_s$.  This rescales all model apparent fluxes by $(D_m/D_s)^2$ and all angular sizes by $D_m/D_s$.
   
   \item Adjust $L_m$ to bring the scaled value of $f_m(12)$ to equal the observed $f_s(12)$; this rescales all angular sizes in proportion to $L_m^{1/2}$.
   The net effect of operations (5) and (6) is to multiply all the original model fluxes by $\epsilon$, the ratio of $f_s(12)$ to the original value of $f_m(12)$, and all angular sizes by the $\sqrt{\epsilon}$.
   
   \item For each observed baseline $BL_s$, look up the correlated flux in the 2D transform of the unscaled model at baseline length $BL_m=BL_s*\sqrt{\epsilon}$ (to account for the rescaled angular size) and multiply this value by $\epsilon$ (to account for the rescaled fluxes).  
   This flux value can now be directly compared to the measured correlated flux at $BL_s$.
   
   \item Repeat these steps for each cloud realization and for all the chosen values of $\theta$ and $\phi$.
   The set of all the correlated fluxes for a given baseline represents the expected distribution of measured fluxes under the assumption of random distributions of these stochastic variables.
\end{enumerate}


\section{Correlation losses}
The atmospheric phase jitter might lead to a reduction of the estimated correlated flux in our measurements. 
To estimate the amount of correlation losses caused during data reduction, we used a similar strategy as explained by \citet{2012SPIE.8445E..1GB}. 
We simulated an observation of a weak target with a known flux to observe the difference between the input and the output flux. 
Since the data used for this work were reduced using EWS 2.0\footnote{EWS is available for download from http://home.strw.leidenuniv.nl/~jaffe/ews/index.html} and the dilution experiment previously reported by \citet{2012SPIE.8445E..1GB} was done using the EWS snapshot version 2012 January 25, we repeated this experiment with the updated version to determine possible changes. 
For our experiment, we took the raw data of a bright target  with known flux, a calibrator in our case, multiplied the input flux by a factor $f<1$ and added artificial noise to the data. 
After performing this process several times, we obtained an estimate of the amount of losses in the flux due to correlation losses. 

\begin{figure}
   \centering
   \includegraphics[width=\hsize]{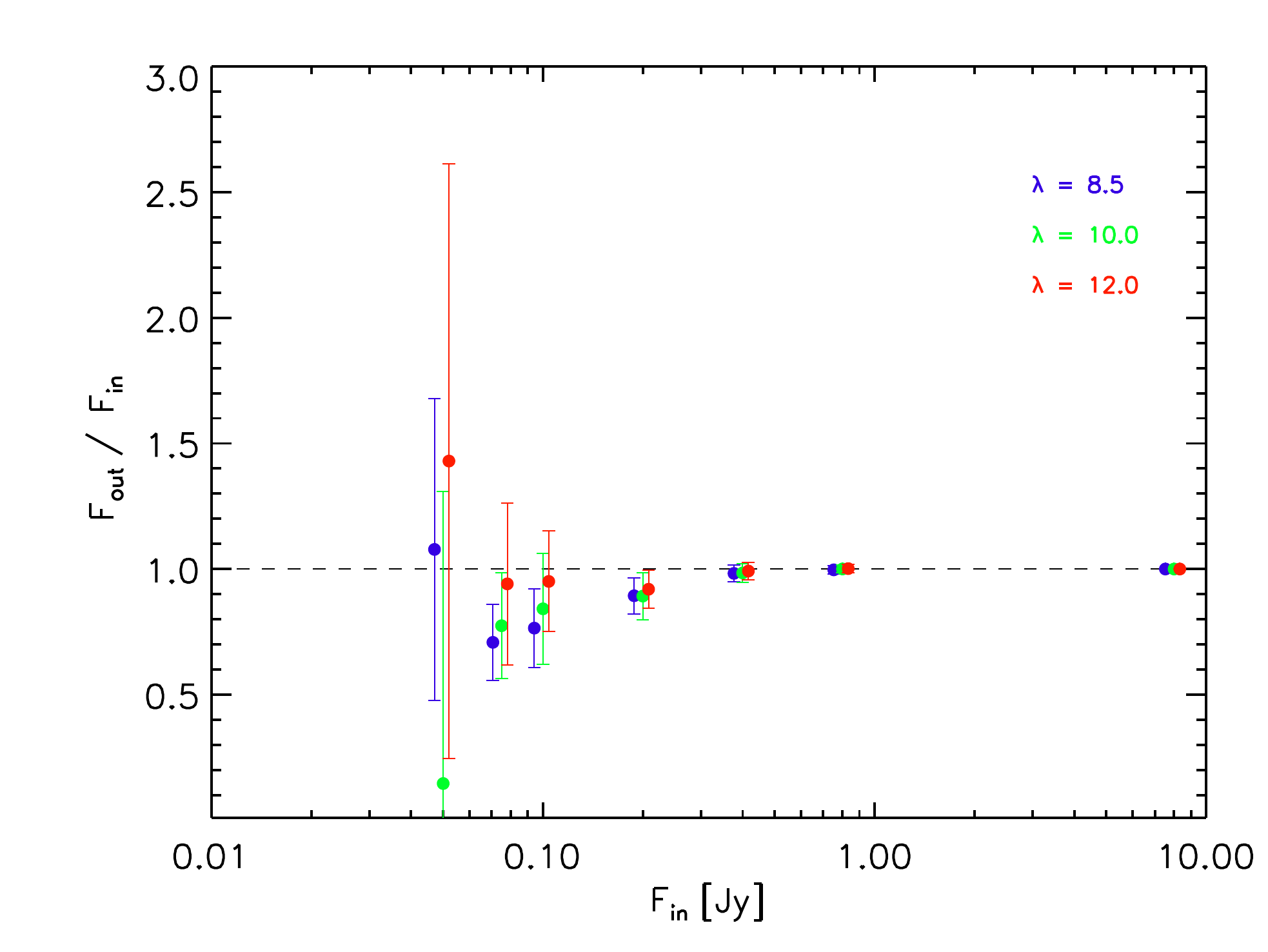}
   \caption{ Average fraction of the recovered flux $F_{out}$ and the input flux $F_{in}$ as a function of the input flux for our dilution experiment at different wavelengths (8.5\,$\mu$m, 10.0\,$\mu$m, and 12.0\,$\mu$m.
   For each value of $F_{in}$ the symbols for the three wavelengths have been shifted slightly for better readability.
   The average values were computed from the output fluxes obtained from 76 calibrators, and the errorbars represent standard deviation of the output fluxes.}
   \label{Figloss}
\end{figure}

We performed this dilution experiment with 76 calibrators observed in different nights and diluted to simulate weak targets with 50\,--\,8000\,mJy of correlated flux.
The losses relative to the highest flux (8\,Jy) were determined at three wavelengths of 8.5, 10.0, and 12.0\,$\mu$m. 
Figure~\ref{Figloss} shows the losses at these three wavelengths, and for the weak sources with fluxes $< 400$\,mJy these losses clearly become more significant at the short wavelength range (closer to $8\,\mu$m), while the losses are moderate above the 
$12\,\mu$m.
We used this information to correct the average correlated fluxes of the sources for this work. 
For the source with high correlation losses we took the average correlated flux obtained after the data reduction and multiplied it by the decorrelation correction factor obtained from our dilution experiment.

\section{Dust sublimation}

\begin{figure}
   \centering
   \begin{minipage}{\hsize}
      \centering
      \includegraphics[width=\hsize]{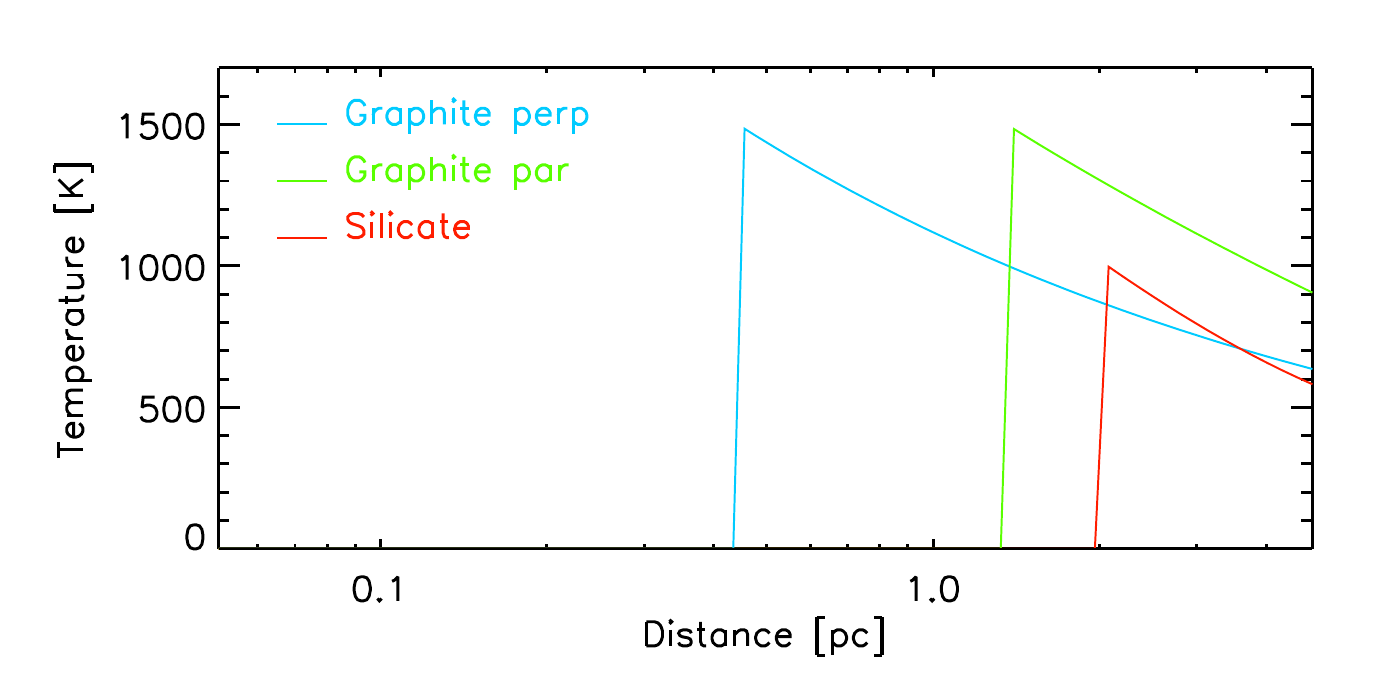}
      \includegraphics[width=\hsize]{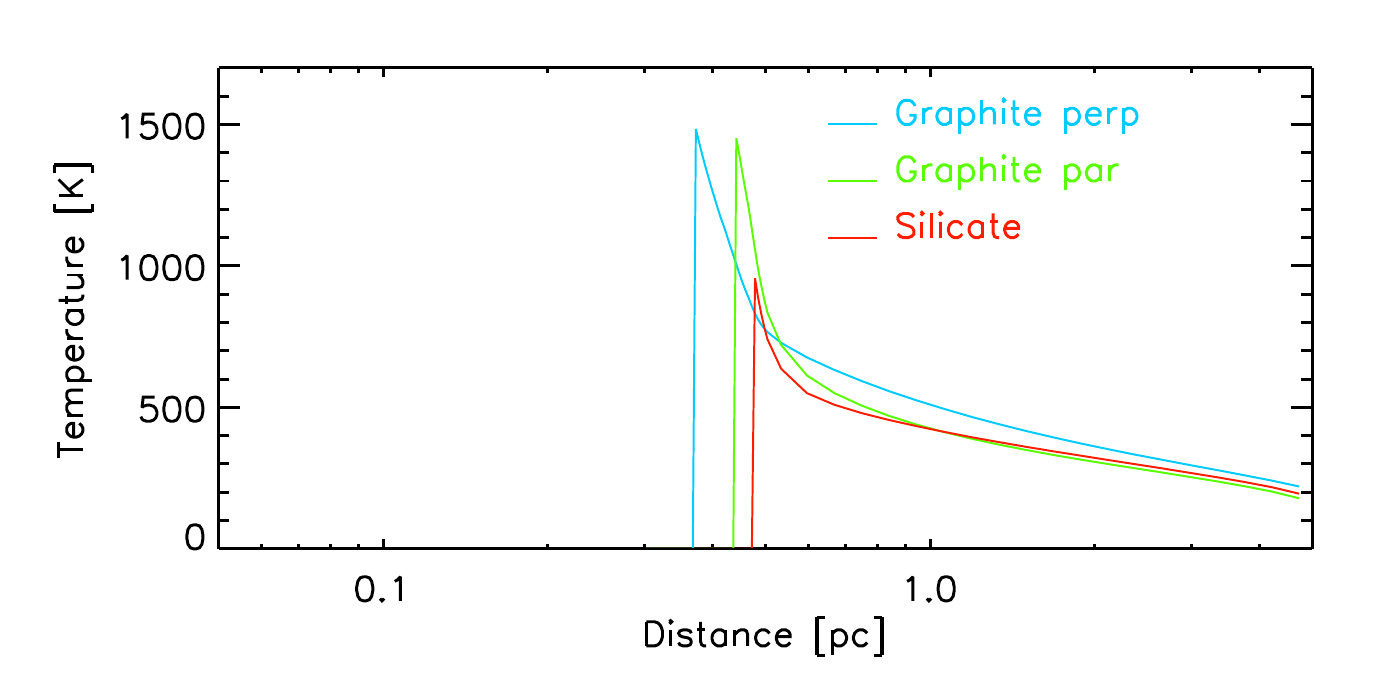}
   \end{minipage}%
   \caption{Temperatures for the three smallest grain sizes of each species in a shell  with a radial optical thickness of $\tau_{9.7}=10^{-4}$ (top) and  $\tau=8$ (bottom).
   The colors indicate the different species of grains: blue - graphites $\perp$, green - graphites $||$, and red - silicates. }
   \label{fig:tempdis}
\end{figure}

\begin{figure}
   \centering
   \includegraphics[width=\hsize]{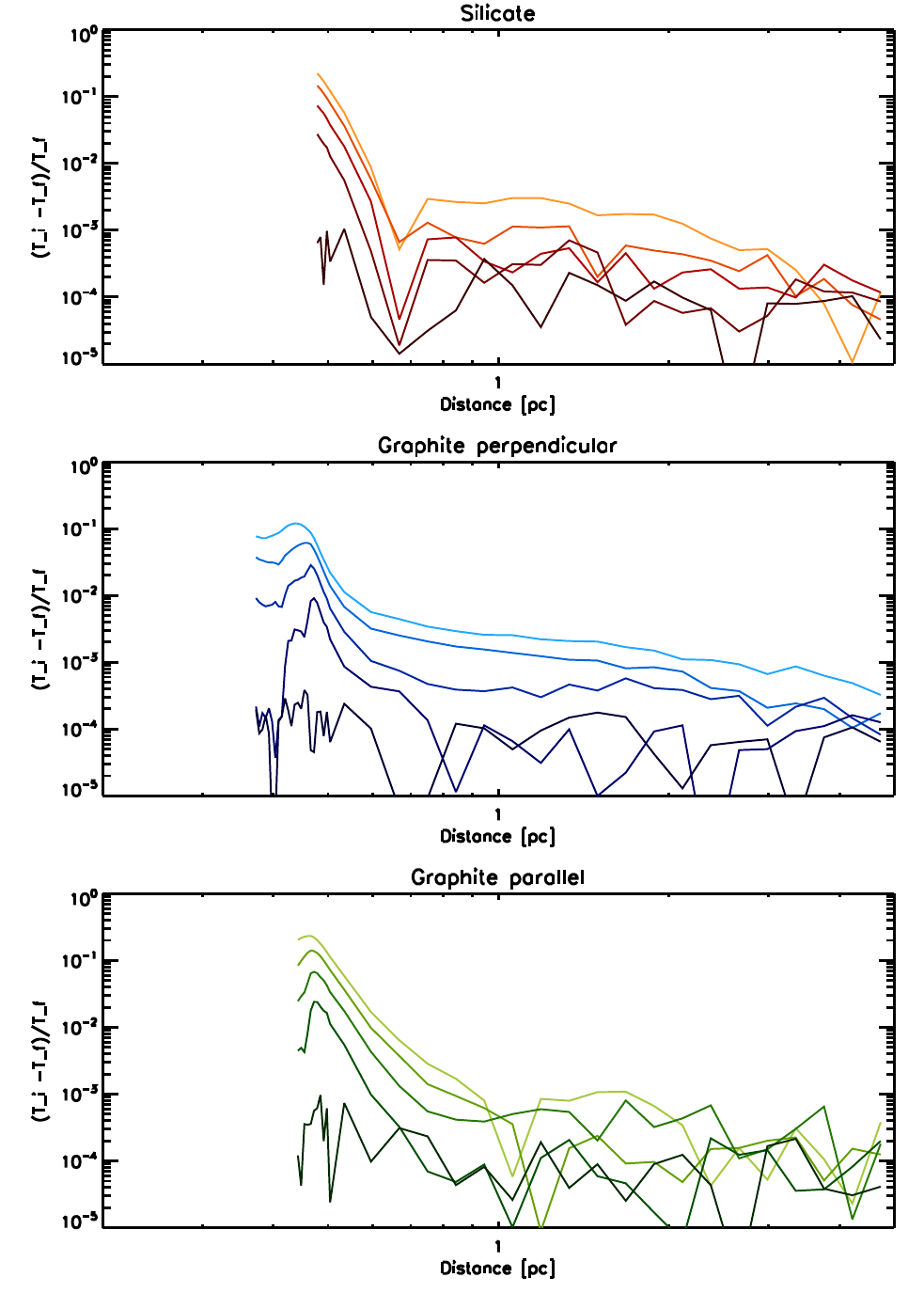}
   \caption{Convergence of the temperature for the three smallest grains. 
   Each line indicate the relative difference after each iteration. 
   The darker the color, the higher the number of iterations.}
   \label{fig:iter}
\end{figure}

The radiative transfer code RADMC-3D does not include an internal computation to account for dust sublimation.
To include dust sublimation in our models, we slightly modified the code. 
Each time a photon package enters a cell, it increases the energy of the cell and thus increases the temperature of the dust of this cell. 
When the dust in the  cells exceeds the dust sublimation temperature, the dust inside the cell is completely removed.
To accurately estimate the temperature of the system using dust sublimation, we performed an iterative process for which the code several times computed the temperatures and removed the corresponding cells. 
When the differences between different iterations were below a certain tolerance (we took a value of 5\,\%), we performed a last computation without removing any dust in the cells to ensure the conservation of energy in the system.
To show how our implementation of the dust sublimation works in different situations we show examples of an optically thick and optically thin case. 
In Fig.~\ref{fig:tempdis} we show the temperature of the three smallest grains of each species for a spherical shell with  $\tau_{9.7}=10^{-4}$ and $\tau_{9.7}=8$, respectively. 
Both shells have an initial inner radius of 0.3\,pc and an outer radius of 5\,pc. We can observe clearly that in the optically thick case, the temperature rises more quickly and the sublimation radius for the three species are closer than in the 
optically thin case. 
We made sure that we had enough cells close to the inner radius to obtain an accurate solution.
In Fig.~\ref{fig:iter}, we show for the optically thick case, the relative differences between each iteration and our final computation.  
After a few iterations, the temperature in every cell reaches a relative difference below our tolerance value.

\section{Acceptance levels}

For our method we used linear transformations computed for each model to remove the mean values, normalize the variance, and remove correlations. 
We expect that if our models agree with the observational data, the final distribution of the data points have a zero mean and variance equal to one. 
To test if this is true for every model, we applied a chi-squared to the sample mean $(\mu)$ and sample variance $(s^2)$ of the normalized measurements, 
\begin{equation}
\chi^2= \left(\frac{\mu}{\sigma_{mean}}\right)^2+\left(\frac{s^2-1}{\sigma_{var}}\right)^2
,\end{equation}
where $\sigma_{mean}$ and $\sigma_{var}$ are the computed variance for the sample mean and the sample variance, respectively.
These two quantities were computed from the distributions of each model. 
Although the distribution produced by our models might not be Gaussian distributions, when we combine all of the measurements to compute the final quantities, the resulting distribution should be more less similar to a Gaussian distribution, according to the central limit theorem.
Therefore, we used the probability values assigned for the $\chi^2$ distribution to derive the level of acceptance for each model.

\begin{figure}
   \centering
   \includegraphics[width=0.7\hsize]{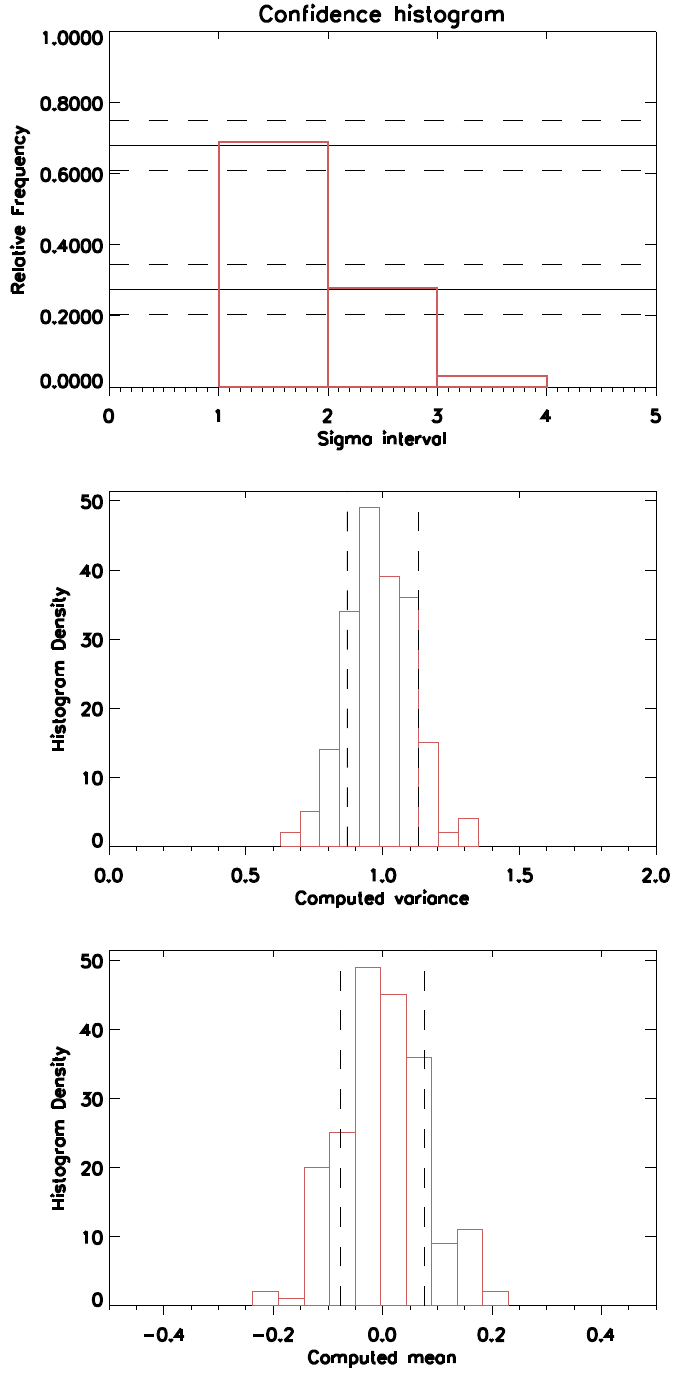}
   \caption{{\it Top)} Histogram showing the frequency of our sample in terms of the  $\sigma$ areas. 
   The lines indicate the 68\,\% and 27.5\,\%, respectively. The dashed lines give the expected uncertainty for 200 experiments. 
   {\it Center and bottom)} Histogram of the sample mean and variance of the normalized measurements for the 200 experiments.}
   \label{fig:self}
\end{figure}

To test if our assumptions are valid, we performed a consistency test.
We took a particular model and created 200 samples with measurements of 20 simulated objects using sparse ($u,v$) coverages and simulating uncertainties of 10\,\%.
For each sample we applied our test with the model that was used to create the simulated samples. 
In Fig~\ref{fig:self} we show the distribution of the sample mean and variance for all the samples. 
We also show that we can safely build our confidence intervals assuming the probabilities for a normal distribution.

\end{appendix}

\end{document}